\documentclass[a4paper,11pt]{article}
\usepackage{jheppub} 
\usepackage{lineno}
\usepackage{float}
\usepackage{subcaption}
\usepackage{gensymb}
\usepackage{comment}
\usepackage{cancel}

\title{\boldmath Photon and neutrino fluxes from spheroidal dwarf galaxies in a decaying DM model}

\author[a]{A. Carrillo-Monteverde}
\author[a,1]{L. López-Lozano \note{Corresponding author.}}
\author[a]{and F. San Juan-Villegas}
\affiliation[a]{\'Area Académica de Matemáticas y Física, Universidad Autónoma del Estado de Hidalgo, Carr. Pachuca-Tulancingo Km. 4.5, C.P. 42184, Pachuca, Hidalgo, México}
\emailAdd{lao\_lopez@uaeh.edu.mx}

\abstract{In this work, we investigate a decaying dark matter scenario and its associated indirect detection signatures. The model consists of a scalar singlet with a lifetime exceeding the age of the Universe. Stability is ensured by a $Z _2$	symmetry imposed on the Lagrangian, allowing decay through a non-minimal gravitational coupling. The decay of dark matter produces Standard Model particles, which subsequently yield products such as gamma rays, neutrinos, and charged particles. We computed the gamma-ray and neutrino fluxes generated by this candidate in the Milky Way and in 14 dwarf spheroidal galaxies, as well as the corresponding expected number of events in selected experiments, using dedicated numerical tools. Results are presented for three benchmark masses and three coupling values consistent with cosmological constraints, showing that the predicted signals can be observable in specific regions of parameter space.
}

\begin{document}
\maketitle
\section{Introduction}
\label{sec:intro}
The nature of dark matter remains one of the most pressing questions in modern astrophysics and particle physics. Despite compelling gravitational evidence for its presence —from galaxy rotation curves and gravitational lensing to the large-scale structure of the Universe— its microscopic properties continue to elude us. Over the past decades, direct detection experiments have pushed sensitivity to unprecedented levels, narrowing the parameter space for weakly interacting massive particles (WIMPs) and other well-motivated candidates \cite{PhysRevLett.131.041003,PhysRevLett.131.041002,PhysRevLett.127.261802}. In particular, the absence of a conclusive signal in direct detection experiments such as XENONnT, LUX-ZEPLIN (LZ), and PandaX has imposed increasingly stringent constraints on the conventional weakly interacting massive particle (WIMP) framework. These null results significantly restrict the viable parameter space of minimal WIMP models and motivate the exploration of alternative scenarios involving suppressed interactions with Standard Model particles, non-minimal dark sectors, or non-thermal production mechanisms.

This persistent absence of direct detection signals is reshaping the landscape of dark matter searches. As a result, indirect detection methods—searching for the secondary products of dark matter annihilation or decay—are gaining renewed prominence \cite{Billard_2022}.

Among these, neutrino telescopes have emerged as uniquely powerful tools \cite{refId0}. By monitoring the faint flux of high-energy neutrinos from astrophysical sources, including the Sun, the Galactic Center, and nearby galaxies, these instruments can probe scenarios that are inaccessible to direct searches. In particular, neutrinos can escape dense environments where other annihilation products are absorbed or scattered, offering a clean and penetrating channel for detecting potential dark matter signatures. Large-scale detectors such as IceCube \cite{Aartsen_2017,IceCube:2017rdn}, ANTARES, and the upcoming KM3NeT \cite{Adrián-Martínez_2016} are now reaching sensitivities capable of testing a broad range of dark matter models, complementing both direct searches and collider experiments.

This work addresses the evolving role of indirect detection, with a focus on photon and neutrino based searches. This analysis has been worked extensively in the context of model with WIMPs without results up today. Therefore, here is considered new sources of information as the dwarf spheroidal galaxies and taking into account the only known DM interaction, this is, gravity. 

Dwarf spheroidal galaxies (dSphs) are among the most promising targets for indirect dark matter detection due to their large dark matter content, low baryonic activity, and reduced astrophysical backgrounds \cite{Strigari_2018}. These properties allow robust determinations of the corresponding J or D-factors, making them ideal laboratories for searching for neutrino fluxes generated by dark matter annihilation or decay. In particular, stacked analyses using IceCube \cite{PhysRevD.108.043001} data have shown that neutrino searches from dSphs can provide competitive constraints on the dark matter annihilation cross section for several leptonic channels.

An appealing alternative to conventional WIMP scenarios is provided by purely gravitational dark matter models, in which the dark sector communicates with the Standard Model exclusively through gravity or Planck-suppressed operators. If DM only interact gravitationally many models has been proposed to parameterize such relations with SM particles. We could mention the models that couple directly Higgs boson with the Ricci scalar as in \cite{Ringwald:1987ui,PhysRevD.98.103521}, models with gravitational dark matter (GDM) \cite{TANG2016402,Garcia:2023qab} and models with pure interaction with gravity \cite{Ema:2018ucl} has been studied. In these frameworks, the observed relic abundance is typically generated through gravitational particle production during reheating or inflation, naturally explaining the absence of direct detection signals while motivating indirect and cosmological probes.

In this paper, the photon and neutrino fluxes coming from dwarf spheroidal galaxies are analyzed in the context of a model with a gravity portal described in \cite{Cata2016,Cata2017,Sun:2020adc}. We show that fluxes coming from the Milky Way and the dSphs have the same order of magnitude and for some scenarios are within the reach of the indirect detection experiments. We focus our attention to the range of masses in the electroweak scale for the main channels at this scale. This paper shows that alternatives explanations to WIMPs are phenomenologically viable.

This paper is organized as follows. In section \ref{model}, the main features of the model describing the DM decay induced by a non minimal gravity coupling with the DM field are reviewed. In section \ref{sec:results}, the photon and neutrino fluxes coming from decaying DM are calculated and discussed for several scenarios. In section \ref{sec:conclusions}, we provide our conclusions.
\section{Decaying DM induced by gravity}
\label{model}
The main feature of decaying models through gravity portals is that the global symmetries that protect DM from decaying are spontaneously broken in curved spaces. The simplest model to show this is the one that extends the SM with a singlet scalar stabilized by a $Z_2$ global symmetry ~\cite{Cata2016}. In the following, we summarize the main features of the model, starting with the classical action in the Jordan frame given by
\begin{equation}
\mathcal{S}=\int d^4x\sqrt{-g}\left[-\frac{R}{2\kappa^2}+\mathcal{L}_\text{SM}+\mathcal{L}_\text{DM}\right],
\end{equation}
where, as usual, $g$ is defined as the determinant of the metric tensor $g_{\mu\nu}$ and $\kappa= \sqrt{8\pi G}$ is the inverse reduced Planck mass. The first term is the Einstein-Hilbert Lagrangian that parameterized the gravitational sector. The SM lagrangian $\mathcal{L}_\text{SM}=\mathcal{T}_F+\mathcal{T}_f+\mathcal{T}_H+\mathcal{L}_Y-\mathcal{V}_H$ considers that SM degrees of freedom live in a curved space that modifies the tensor products with the presence of a non flat metric. The kinetic terms of a gauge boson and the standard fundamental interactions in the scalar sector are written as
\begin{align}
	\mathcal{T}_F&= -\frac{1}{4}g^{\mu\nu}g^{\lambda\rho}F^a_{\mu\lambda}F^a_{\nu\rho},\\
	\mathcal{T}_H& =g^{\mu\nu}(D_\mu H)^\dagger(D_\nu H).
\end{align}
In order to consider the presence of gravity in the gauge interactions, the covariant derivative is modified defining $\cancel{\nabla}=\gamma^a e^\mu_aa \nabla_\mu$ such that the fermion sector is given by 
\begin{equation}
	\mathcal{T}_f=\frac{i}{2}\bar{f}\overleftrightarrow{\cancel{\nabla}}f,
\end{equation}
where $\nabla_\mu=D_\mu-(i/4)e^b_\nu(\partial_\mu e^{\nu c})\sigma_{bc}$, being $D_\mu$ the usual covariant derivative and $e^{\mu c}$ a vierbein obeying $\gamma^\mu=e^\mu_a \gamma^a$.
In order to parameterize the gravity portal, the simplest linear interaction with the Ricci scalar is considered through a function invariant under Lorentz and gauge symmetries that depends on the DM fields. The proposed function can be cast as $F(\varphi, X)$, where $\varphi$ is the DM scalar and $X$ are possible extra non standard particles. Thus, the interaction of DM with the Ricci scalar can be introduced with
\begin{equation}
	\mathcal{L}_\xi=-\xi R F(\varphi,X),
\end{equation}
where $\xi$ is the coupling constant and $F(\varphi,X)$ the real function containing the main contribution of the DM fields. The introduction of these interactions allows us to define the function of the DM fields as
\begin{equation}
    \Omega^2(\varphi,X)=1+2\kappa^2 \xi F(\varphi,X),
\end{equation}
that acts as a conformal transformation $\tilde{g_{\mu\nu}}=\Omega^2(\varphi,X)g^{\mu\nu}$ allowing us to rewrite the action in the equivalent Einstein frame,
\begin{equation}\label{eq:Einstein_action}
    \mathcal{S}=\int d^4x\sqrt{-\tilde{g}}\left[-\frac{R}{2\kappa^2}+\frac{3}{\kappa^2}\frac{\tilde{\nabla}_\mu\Omega\tilde{\nabla}^\mu\Omega  }{\Omega^2}+\tilde{\mathcal{L}}_\text{SM}+\tilde{\mathcal{L}}_\text{DM}\right].
\end{equation}
with the lagrangians taking the following form
\begin{align}
    \tilde{\mathcal{L}}_{SM}&=\tilde{\mathcal{T}}_F+\Omega^{-3}\tilde{\mathcal{T}}_f+\Omega^{-2}\tilde{\mathcal{T}}_H+\Omega^{-4}(\mathcal{L}_\text{Y}-\mathcal{V}_\text{H})\\
    \tilde{\mathcal{L}}_\text{DM}&=\Omega^{-2}\tilde{\mathcal{T}}_\varphi - \Omega^{-4} V(\varphi,X).\label{eq:l_dm}
\end{align}
The equation \eqref{eq:l_dm} yields the coupling of the SM fields with the DM candidates through a Taylor expansion with respect to the coupling $\xi$. The detailed expression of the Feynman rules can be seen in the paper of O.Catà et. al. ~\cite{Cata2017}. In this work, we focus our attention on the channels that contribute abundantly in the electroweak scale because they comprehend the most reliable detection possibilities ~\cite{Sun:2020adc}. The action in \eqref{eq:Einstein_action} allows us to introduce several candidates for the DM fields as in ~\cite{Cata2017} and we focus our attention to the simplest extension of the SM containing a real scalar field as the DM candidate. In this sense, the model is reduced to:
\begin{equation}
    \mathcal{L}_\text{DM}=\frac{1}{2}g_{\mu\nu}
\partial^\mu \phi\partial^\nu\phi-V(\phi,H),
\end{equation}
where $\phi$ is a real scalar field odd under the global $Z_2$ symmetry and the potential containing the kinetic terms and the portal with the Higgs doublet,
\begin{equation}
    V(\phi,H)=\frac{1}{2}\mu_\phi^2\phi^2+\frac{1}{4!}\lambda_\phi\phi^4+\frac{1}{2}\lambda_{\phi H}\phi^2(H^\dagger H).
\end{equation}
The strength of the portal is determined by the parameter $\lambda_{\phi H}$ and is bounded by the measurement of the relic density via thermal freeze-out. Thus, for simplicity, the contribution of the DM in this model is given by $F(\phi,X)\sim \xi\phi$, a linear dependence on the DM candidate. Hence, the Weyl transformation on the lagrangian to obtain the Einstein representation of the action is given by the factor
\begin{equation}
    \Omega^2(\phi)=1+2\xi M \kappa^2\phi,
\end{equation}
with $M$ a mass scale that can be closed to the Planck mass $M_P$.
After the Taylor expansion in powers of $\phi$ up to linear terms, the tree level effective interactions with the SM fields are obtained:
\begin{equation}
    \tilde{\mathcal{L}}_{\text{SM},\phi}=-2\xi M \kappa^2\phi\left[\frac{3}{2}\tilde{T}_f+\tilde{T}_H+2(\mathcal{L}_\text{Y}-\mathcal{V}_H)\right].
\end{equation}
As discussed in ~\cite{Cata2016,Cata2017}, the scale of the kinematically favored decay modes of $\phi\to hh,WW,ZZ$ and $\phi\to ff$ goes as $m_\phi^3$ and $m_f^2m_\phi$ respectively. At the electroweak scale, these channels represent the main contribution to the neutrino fluxes coming from the decaying DM, thus in this work, only these channels are taken into account. The Feynman rules of these channels are shown in Table ~\ref{tab:feynman_rules}.

\begin{table}[h]
    \centering
    \begin{tabular}{cc}
        \hline
        \hline
         Channel & Feynman rules\\
         \hline
         $\phi\to \bar{f}_if_i$& $i\xi\kappa^2 m_{f_i}$ \\
         $\phi\to hh$& $2i\xi\kappa^2(p_1\cdot p_2+2m_h^2)$\\
         $\phi\to V_\mu V_\nu$ & $-2i\kappa^2 M m_V^2\eta_{\mu\nu}$\\
         \hline
         \hline
    \end{tabular}
    \caption{Feynman rules for the main contribution to neutrino fluxes at the electroweak scale. The field $V_\mu$ represents the gauge boson $W$ and $Z$. In the fermion field $f_i$, the index represents the generation.}
    \label{tab:feynman_rules}
\end{table}
\section{Photon and neutrino fluxes from decaying DM}
\label{sec:results}
The Milky Way and dwarf spheroidal galaxies (dSphs) are among the most relevant astrophysical targets in indirect searches for dark matter (DM), due to their proximity and large DM content. In this context, the main goal is to identify signals produced by DM decay or annihilation, typically in the form of gamma rays, neutrinos, or charged particles.

In the case of decaying DM, gamma-ray and neutrino fluxes can be computed as~\cite{Cirelli}:
\begin{equation}
     \frac{d \Phi_{X}}{d\Omega dE}=\frac{r_{s}}{4\pi}\frac{\rho_{0}}{M_{DM}}D\sum_{f}\Gamma_f\frac{dN_{X}^f}{dE},
    \label{E:fluxes}
\end{equation}
where $dN_{X}^f/dE$ is the photon/neutrino energy spectrum for the final state $f$, $\Gamma_f$ is the corresponding branching ratio, and $D$ is the astrophysical factor given by
\begin{equation}
        D=\int_{l.o.s}\frac{ds}{r_{s}}\left( \frac{\rho(r(s,\theta))}{\rho_{0}}\right),
        \label{E:D-factor}
\end{equation}
with $r$ the distance from the center of the halo and $\theta$ the observation angle.

In this work, we analyzed 14 different dSphs, as shown in Table ~\ref{tab:infor}, where the main information about them is presented. 

\begin{table}
    \centering
    \begin{tabular}{c|c|c|c|c|c|c}
    \hline
        Name & l ($^\circ$) & b ($^\circ$) & d (kpc) & R$_\Delta$ (kpc) & $\rho_s$ (M$_\odot$/kpc$^3$) & $r_s$ (kpc) \\
    \hline
        Bootes I & 358.08 & 69.62 & 66 & 0.541 & 18197008 & 6.309 \\
        Canes Venatici I & 74.31 & 79.82 & 218 & 2.016 & 13489628 & 2.290 \\
        Canes Venatici II & 113.58 & 82.70 & 160 & 0.363 & 34673685 & 8.128 \\
        Coma Berenice & 241.89 & 83.61 & 44 & 0.238 & 102329299 & 5.128 \\
        Draco & 86.37 & 37.42 & 76 & 1.724 & 18197008 & 3.715 \\
        Hercules & 28.73 & 36.87 & 132 & 0.645 & 23988329 & 0.851 \\
        Leo I & 225.99 & 49.11 & 254 & 1.994 & 6606934 & 6.309 \\
        Leo II & 220.17 & 67.23 & 233 & 0.935 & 120226443 & 0.776 \\
        Leo IV & 265.44 & 56.51 & 154 & 0.430 & 15848931 & 0.870 \\
        Segue I & 220.48 & 50.43 & 23 & 0.140 & 87096358 & 1.819 \\
        Sextans & 243.50 & 42.27 & 86 & 2.551 & 4073802 & 6.165 \\
        Ursa Major I & 159.43 & 54.41 & 97 & 0.897 & 14454397 & 3.162 \\
        Ursa Major II & 152.46 & 37.44 & 32 & 0.240 & 74131024 & 4.265 \\
        Triangulum II & 33.32 & 36.18 & 30 & 10 & 30052000 & 0.1 \\
    \hline
    \end{tabular}
    \caption{Information about the dSphs studied for this work. Columns 2 and 3 correspond to the galactic coordinates, $d$ is the distance to the object, R$_\Delta$ is the outer bound of the halo and $\rho_s$, $r_s$ are parameters for the density profile ~\cite{Albert, lat}.}
    \label{tab:infor}
\end{table}

The information in Table ~\ref{tab:infor} is needed to run \texttt{CLUMPY} ~\cite{Charbonnier_2012, Bonnivard:2015pia, Hutten:2018aix}, and compute the fluxes of indirect signals produced in these objects. 

\begin{figure}
    \centering
    \includegraphics[width=0.5\linewidth]{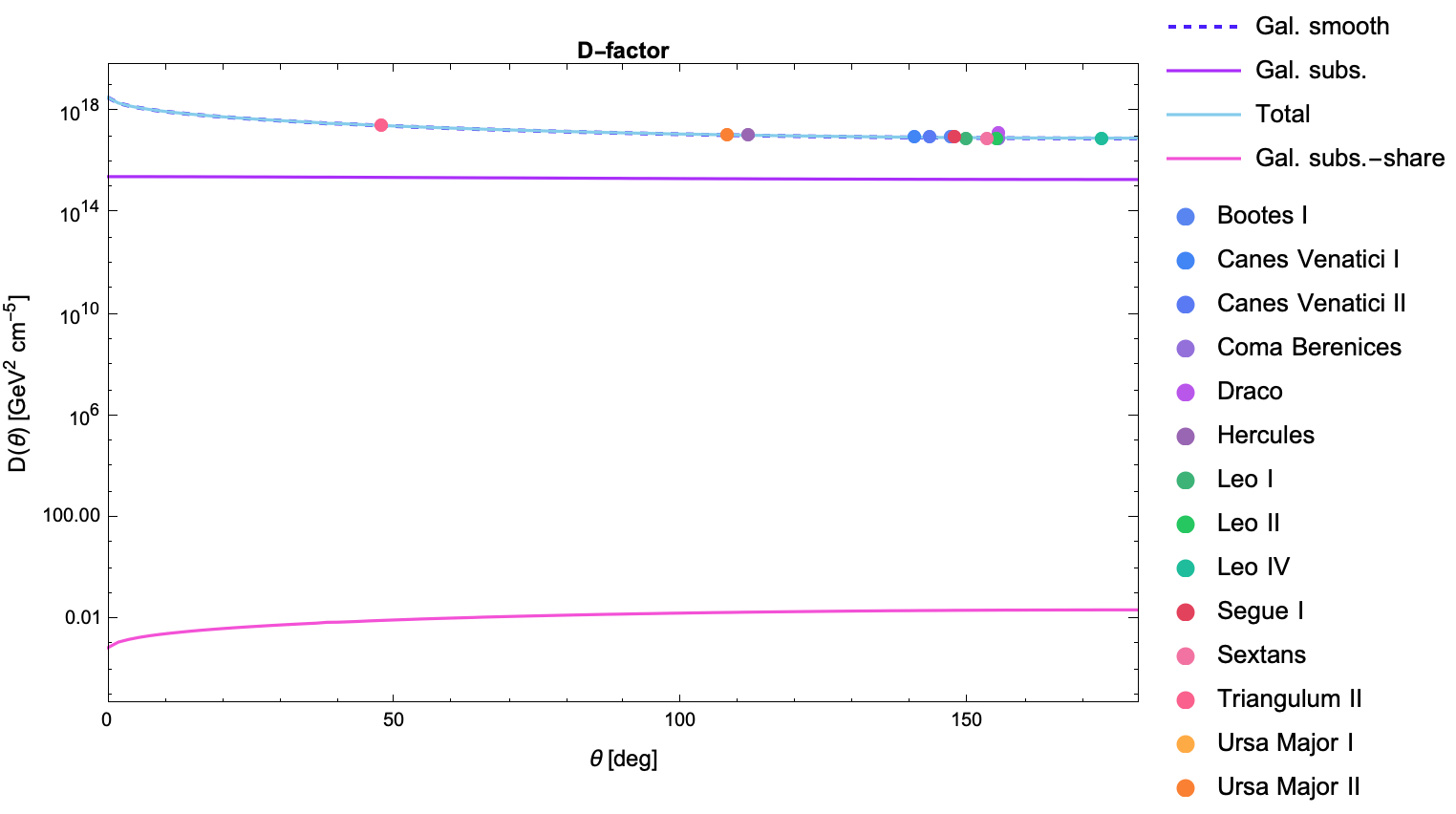}
    \caption{Comparative of D-factors from the Milky Way and some dSphs galaxies. The solid blue line corresponds to the D-factor of the Milky Way and the colored dots are for the dSpshs galaxies. The order of the magnitude is similar, so the density of DM in our galaxy does not differ much from the dSphs galaxies.}
    \label{fig:D-factor-Comp}
\end{figure}

First, from \eqref{E:fluxes}, we observe that the D-factor is important to provide information about the distribution of DM. We computed the D-factor, using \texttt{CLUMPY}, and made a comparison between the objects studied in this work. The results obtained are shown in Figure ~\ref{fig:D-factor-Comp}.

The total number of events, $\mu_s$, that can be detected in a neutrino telescope is calculated using the equation ~\cite{Chianese:2019kyl}:

\begin{equation}
    \mu_s=T \sum_{\alpha} \int d\Omega \int dE_{\nu}A_{\nu_\alpha}^{eff} (\Omega, E_\nu) \Phi_{\nu_\alpha}(E_\nu),
    \label{E:Events}
\end{equation}

where $T$ is the exposure time, $A_{\nu_\alpha}^{eff} (\Omega, E_\nu) $ is the effective area for a specific flavor $\alpha$ that depends on the observation angle and the neutrino energy; and  $\Phi_{\nu_\alpha}(E_\nu)$ is the flux of the flavor $\alpha$. In the case of gamma rays, we can use the same equation but there is not dependence on the flavor.

\subsection{Milky Way}
\label{sub:milky}

Our Galaxy is one of the objects commonly used in the indirect detection of DM due to the high density in its halo. This halo is expected to have a density of DM that decreases with the distance from the center of the galaxy, and due to this, the center of the Milky Way is also a significant source of gamma ray production ~\cite{Fornasa}.

To study the distribution of DM in our Galaxy several density profiles can be used; their behavior is different in the inner part of the galactic halo but similar in the ratio of some kiloparsecs, for example, at the distance of our Sun. The DM signals produced near the center of the galaxy are more sensitive than those produced near Earth ~\cite{Cirelli}. 

\subsubsection{Gamma Rays}

With the value of the D-factor and using equation \eqref{E:fluxes}, we obtained the fluxes of gamma rays produced in the halo that involves our galaxy. 

\begin{figure}
    \centering
    \begin{minipage}[b]{0.305\textwidth}
        \centering
        \includegraphics[width=\textwidth]{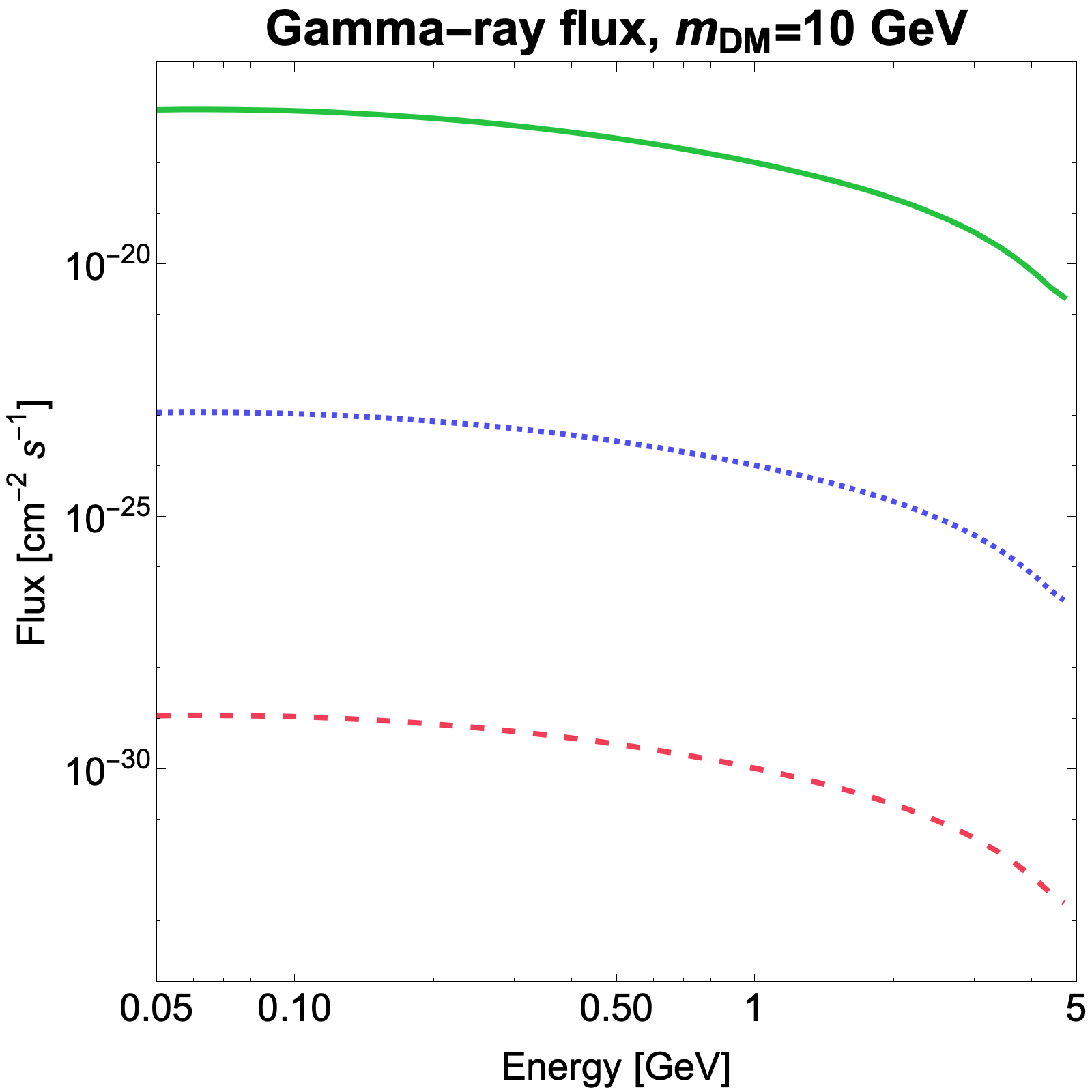}
        \subcaption{}
    \end{minipage}
    \begin{minipage}[b]{0.305\textwidth}
        \centering
        \includegraphics[width=\textwidth]{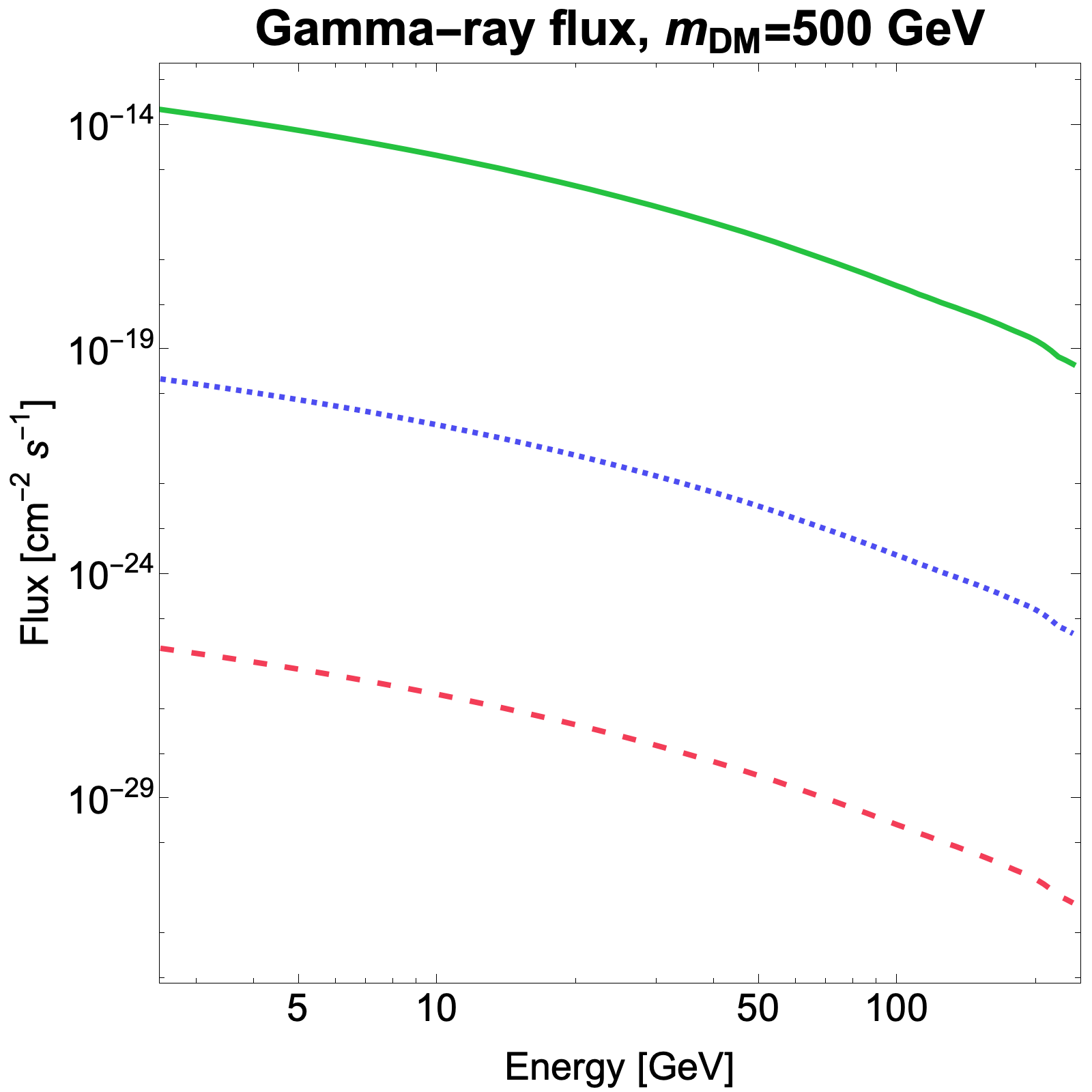}
        \subcaption{}
    \end{minipage}
    \begin{minipage}[b]{0.37\textwidth}
        \centering
        \includegraphics[width=\textwidth]{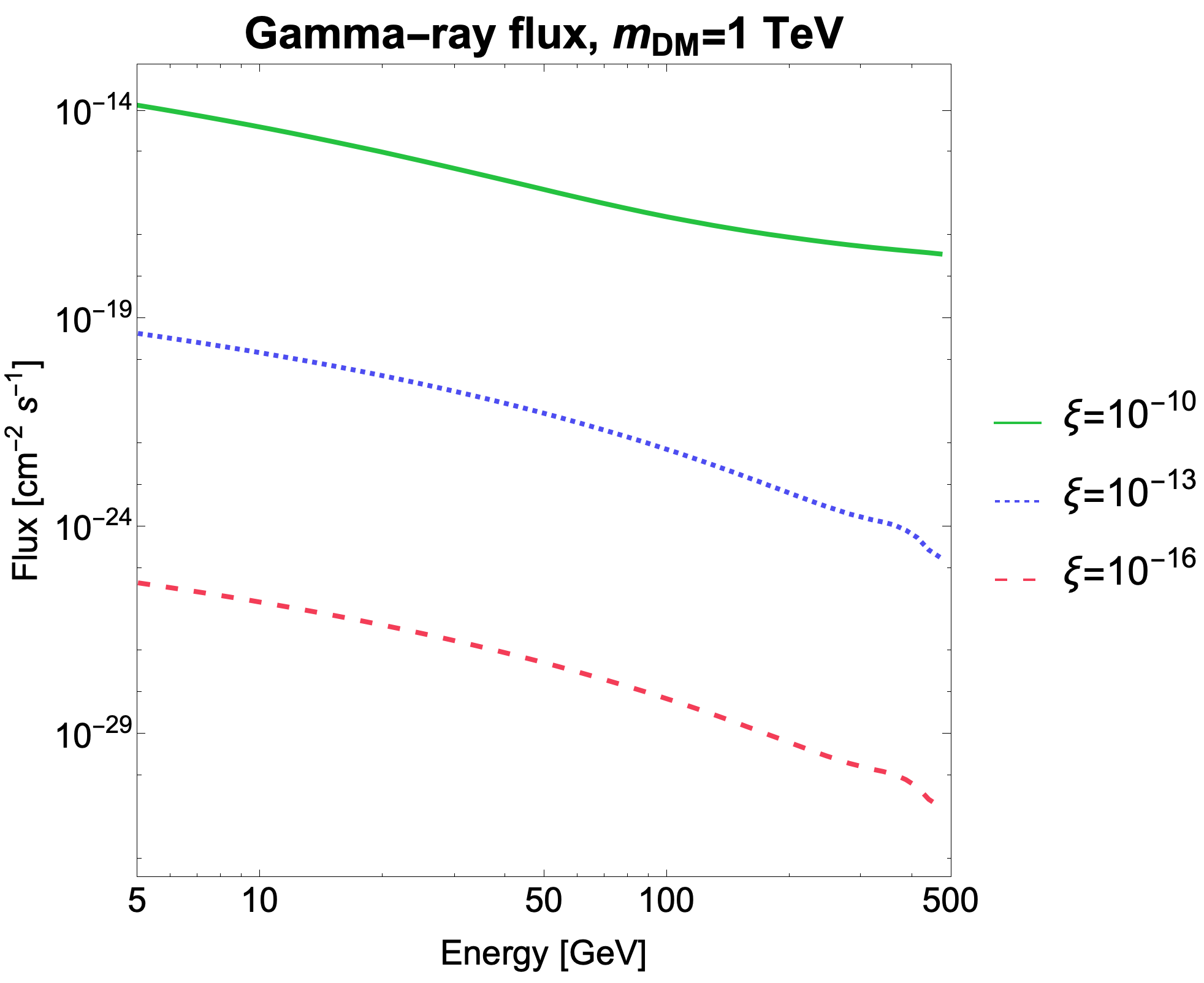}
        \subcaption{}
    \end{minipage}
    \caption{Flux of gamma rays varying the mass and the coupling parameter. In these plots, the solid green line corresponds to flux considering $\xi=10^{-10}$, the solid blue line is for $\xi=10^{-13}$ and the dotted red line corresponds to $\xi=10{-16}$. The mass for the candidate is fixed in each graph, in figure a) we consider a mass of 10 GeV, in b) 500 GeV and for c) is 1 TeV.}
    \label{fig:Gamma-flux-MW}
\end{figure}

Figure ~\ref{fig:Gamma-flux-MW} shows the differential flux of gamma rays. For the first graph, the mass value is fixed, 10 GeV, and $\xi$ is the parameter that changes, in the second graph the mass is 500 GeV and in the third one is 1 TeV. To obtain the data for the graphs the module \texttt{g2} in \texttt{CLUMPY} was run, with the following information from the model: the BRs of the decaying channels, lifetime and mass of DM and energy for the differential flux. We had to run the module for each value in the range of the mass on the x-axis. The highest value for the flux is the one with a DM mass of 1 TeV and $\xi=10^{-10}$, while the lowest value is considering a mass of 10 GeV and $\xi=10^{-16}$. Between the first and second graphs there is a difference of six orders of magnitude, but between the second and third ones the difference is only close to one order of magnitude for the case of the minimum value of the coupling parameter. 

Using the values of the differential flux, the number of events that can be detected on Earth can be estimated, considering a detector with an effective area of one squared kilometer and a year of exposure. 

\begin{figure}
    \centering
    \begin{minipage}[b]{0.305\textwidth}
        \centering
        \includegraphics[width=\textwidth]{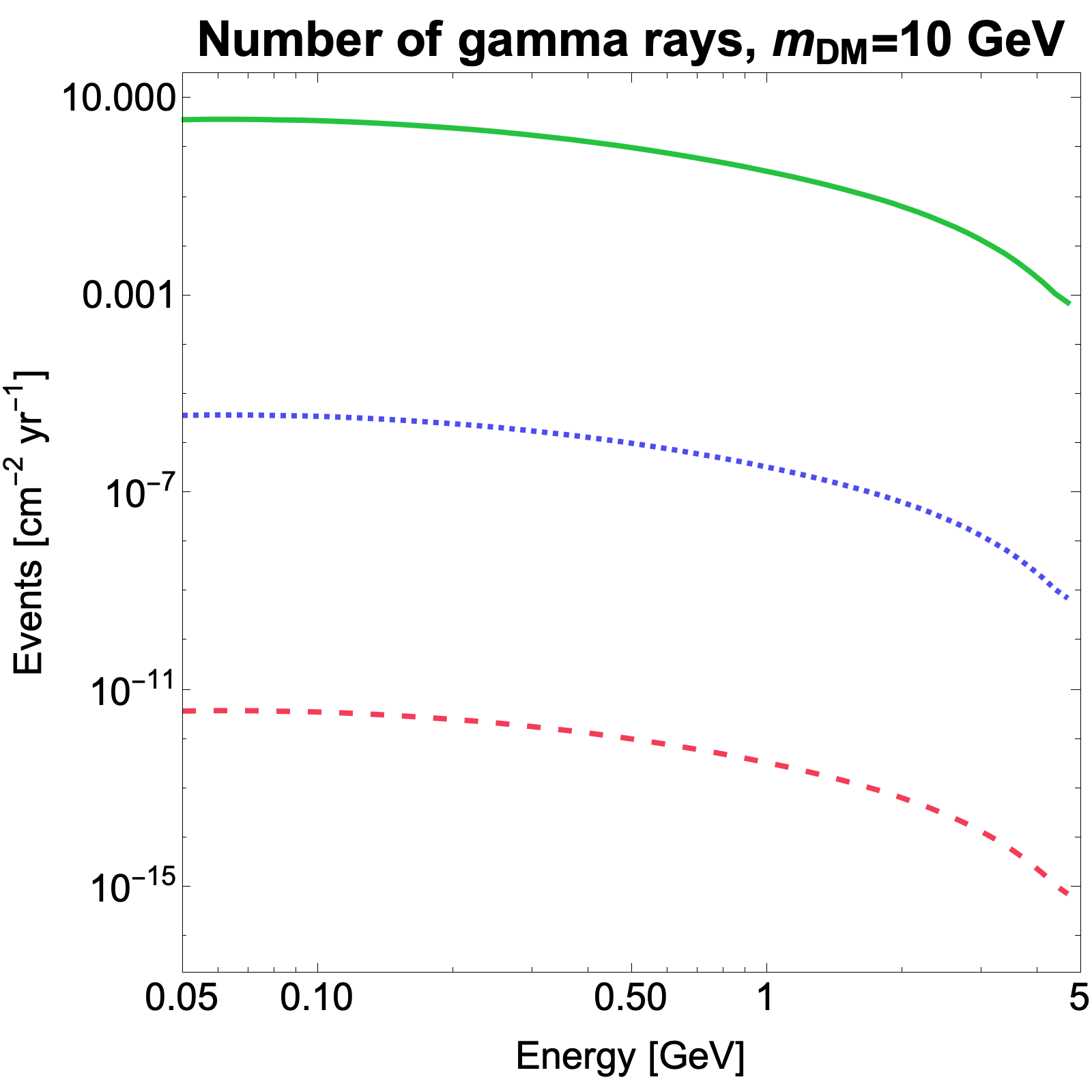}
        \subcaption{}
    \end{minipage}
    \begin{minipage}[b]{0.305\textwidth}
        \centering
        \includegraphics[width=\textwidth]{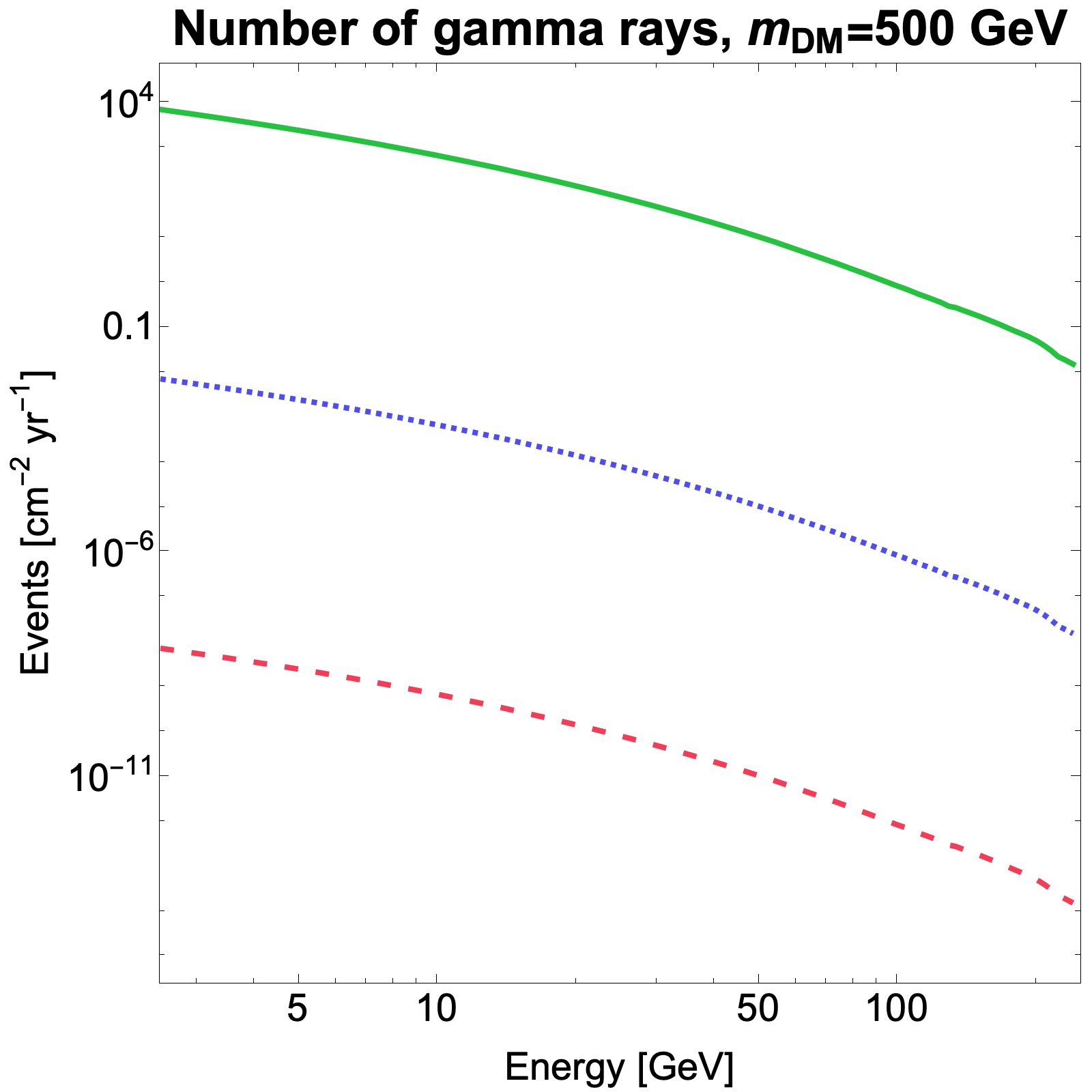}
        \subcaption{}
    \end{minipage}
    \begin{minipage}[b]{0.37\textwidth}
        \centering
        \includegraphics[width=\textwidth]{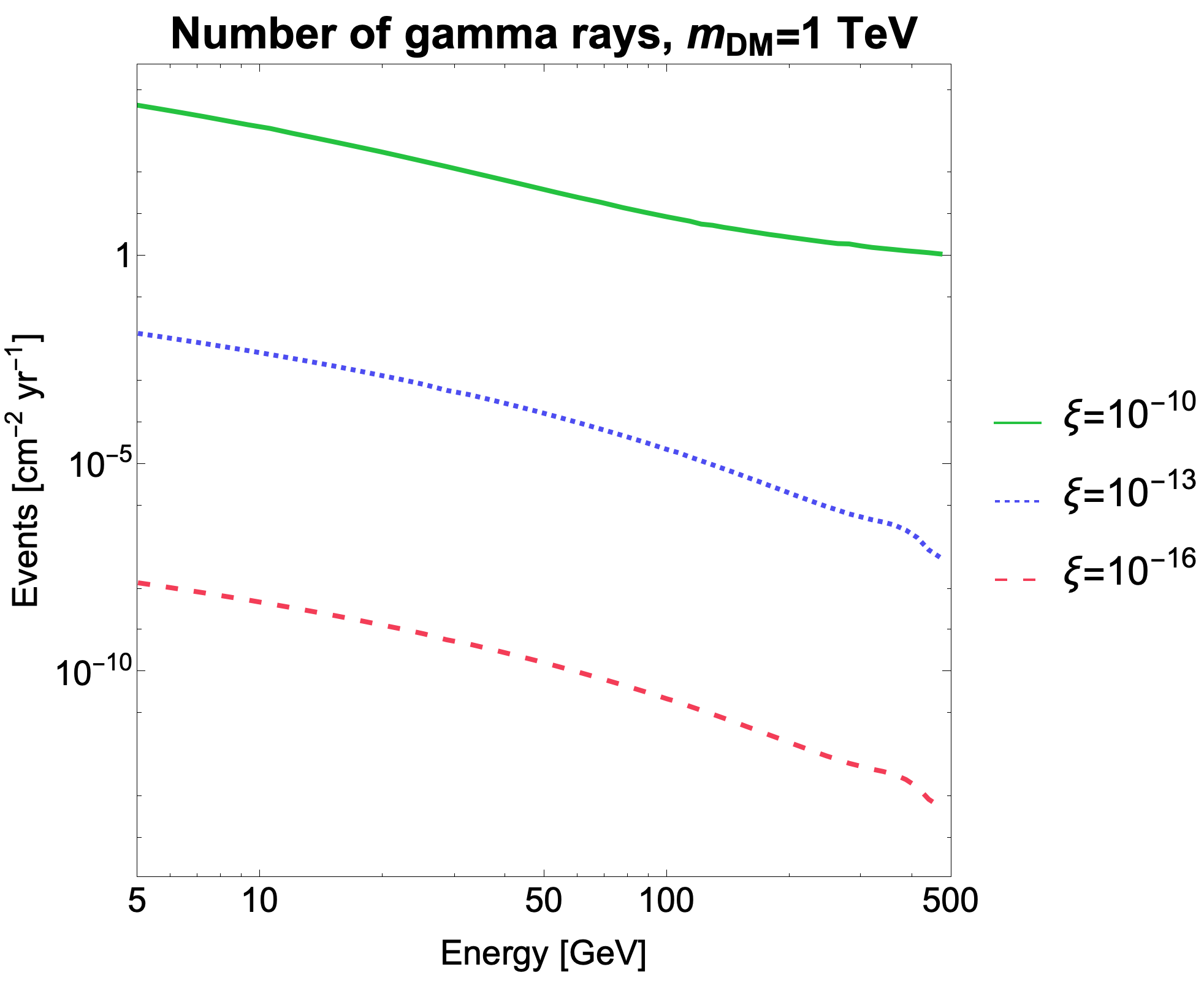}
        \subcaption{}
    \end{minipage}
    \caption{Number of gamma rays varying the mass and the coupling parameter. In these plots, the solid green line corresponds to the number of events for $\xi=10^{-10}$, the solid blue line is for $\xi=10^{-13}$ and the dotted red line corresponds to $\xi=10{-16}$. The mass for the candidate is fixed in each plot, in figure a) we consider a mass of 10 GeV, in b) 500 GeV and for c) is 1 TeV.}
    \label{fig:Gamma-eve-MW}
\end{figure}

As expected and with concordance in Figure ~\ref{fig:Gamma-flux-MW}, the best result for the detection of this signal is the case with a mass of 1 TeV and $\xi=10^{-10}$. Although the values are really close to the mass of 500 GeV. We can observe that the behavior is almost the same in all the plots; for the small values of mass, the flux and number of events are greater than for the largest values of energy. 

\subsubsection{Neutrinos}

\begin{figure}
    \centering
    \begin{minipage}[b]{0.305\textwidth}
        \centering
        \includegraphics[width=\textwidth]{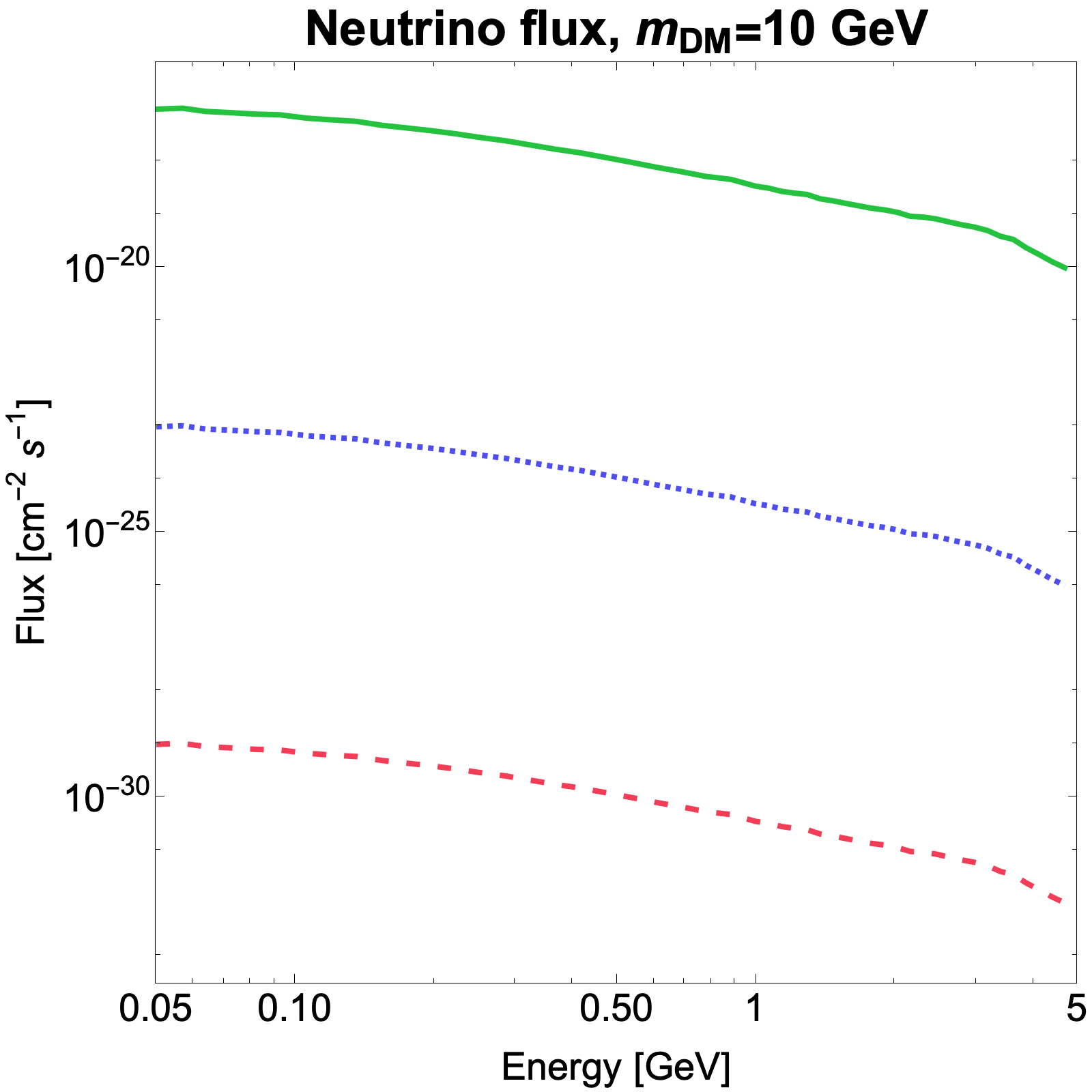}
        \subcaption{}
    \end{minipage}
    \begin{minipage}[b]{0.305\textwidth}
        \centering
        \includegraphics[width=\textwidth]{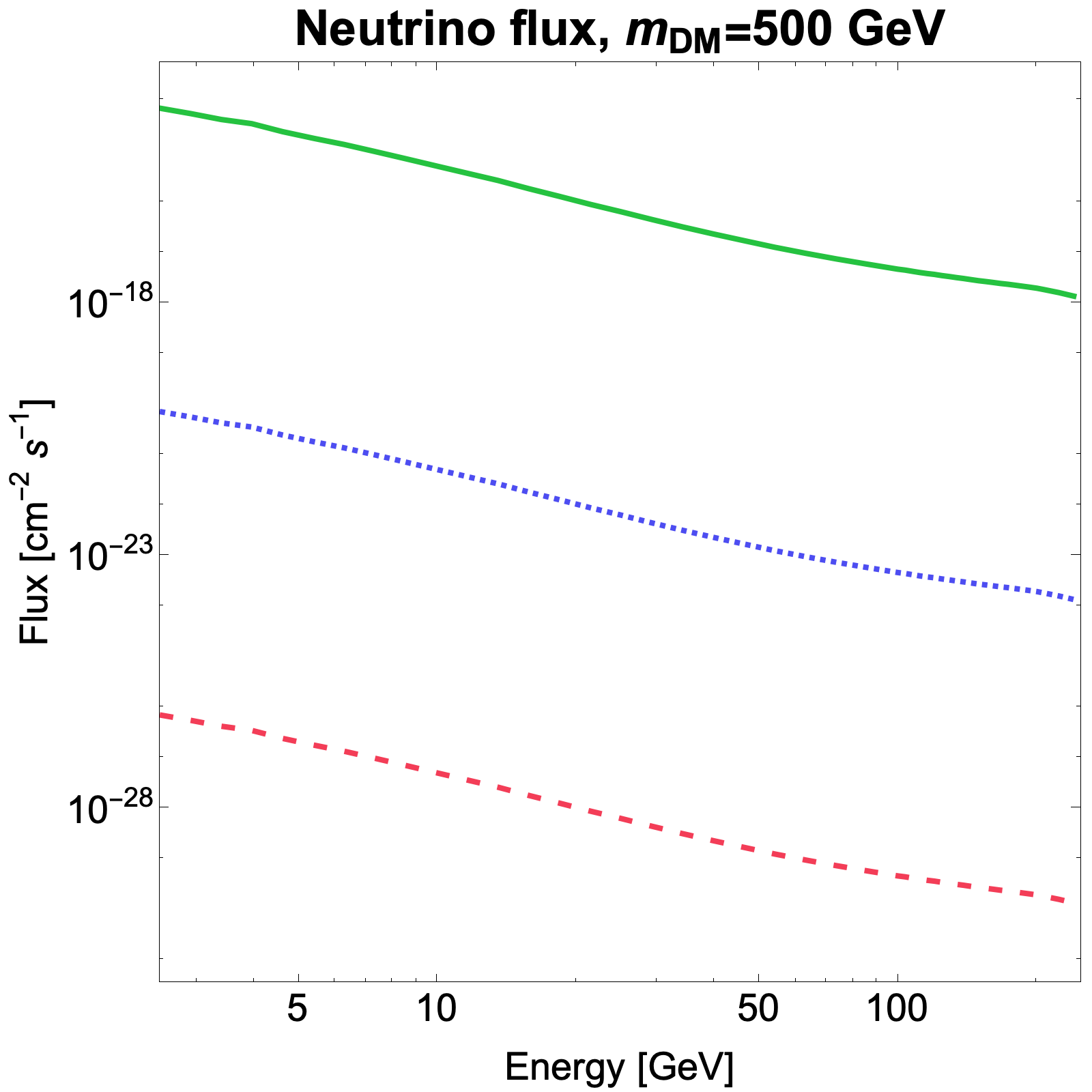}
        \subcaption{}
    \end{minipage}
    \begin{minipage}[b]{0.37\textwidth}
        \centering
        \includegraphics[width=\textwidth]{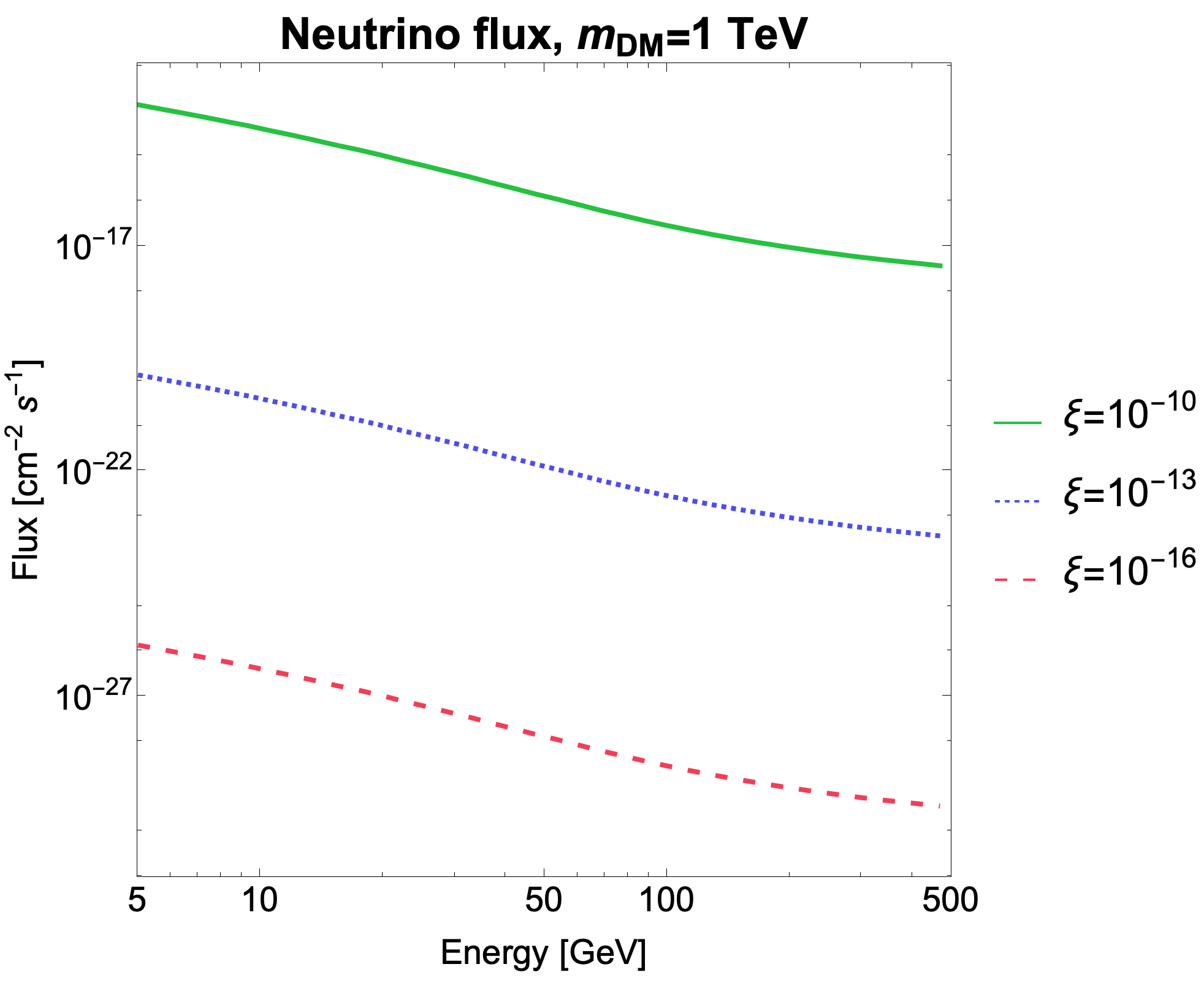}
        \subcaption{}
    \end{minipage}
    \caption{Flux of neutrinos varying the mass and the coupling parameter. In these plots, the solid green line corresponds to flux of muon neutrinos that arrives to Earth, considering $\xi=10^{-10}$, the solid blue line is for $\xi=10^{-13}$ and the dotted red line corresponds to $\xi=10{-16}$. The mass for the candidate is fixed in each plot, in Figure a) we consider a mass of 10 GeV, in b) 500 GeV and for c) is 1 TeV.}
    \label{fig:Nu-flux-MW}
\end{figure}

With the same module used in the case of gamma rays, we also obtained the corresponding results for neutrinos. For the flux of neutrinos produced in the halo of our Galaxy, the results are shown in Figure ~\ref{fig:Nu-flux-MW}. In this case, as the fluxes are computed at production, the oscillation of neutrinos was considered to compute the total flux that arrives to Earth, and the muon neutrino was chosen as the final state. With the data obtained from \texttt{CLUMPY}, we processed the files in \texttt{PYTHON} with a program that multiplies the flux in the source by the probability of oscillation, as the flux for each flavor in the source was computed.

\begin{equation}
    \text{Total flux}= \text{flux}(\nu_{\mu}\rightarrow\nu_{\mu})+ \text{flux}(\nu_{e}\rightarrow\nu_{\mu})+\text{flux}(\nu_{\tau}\rightarrow\nu_{\mu})
\end{equation}

With the flux, we computed the number of neutrinos that can be detected in Earth with a detector with one squared kilometer of effective area and one year data. The neutrino flavor considered is the muon type.

\begin{figure}
    \centering
    \begin{minipage}[b]{0.305\textwidth}
        \centering
        \includegraphics[width=\textwidth]{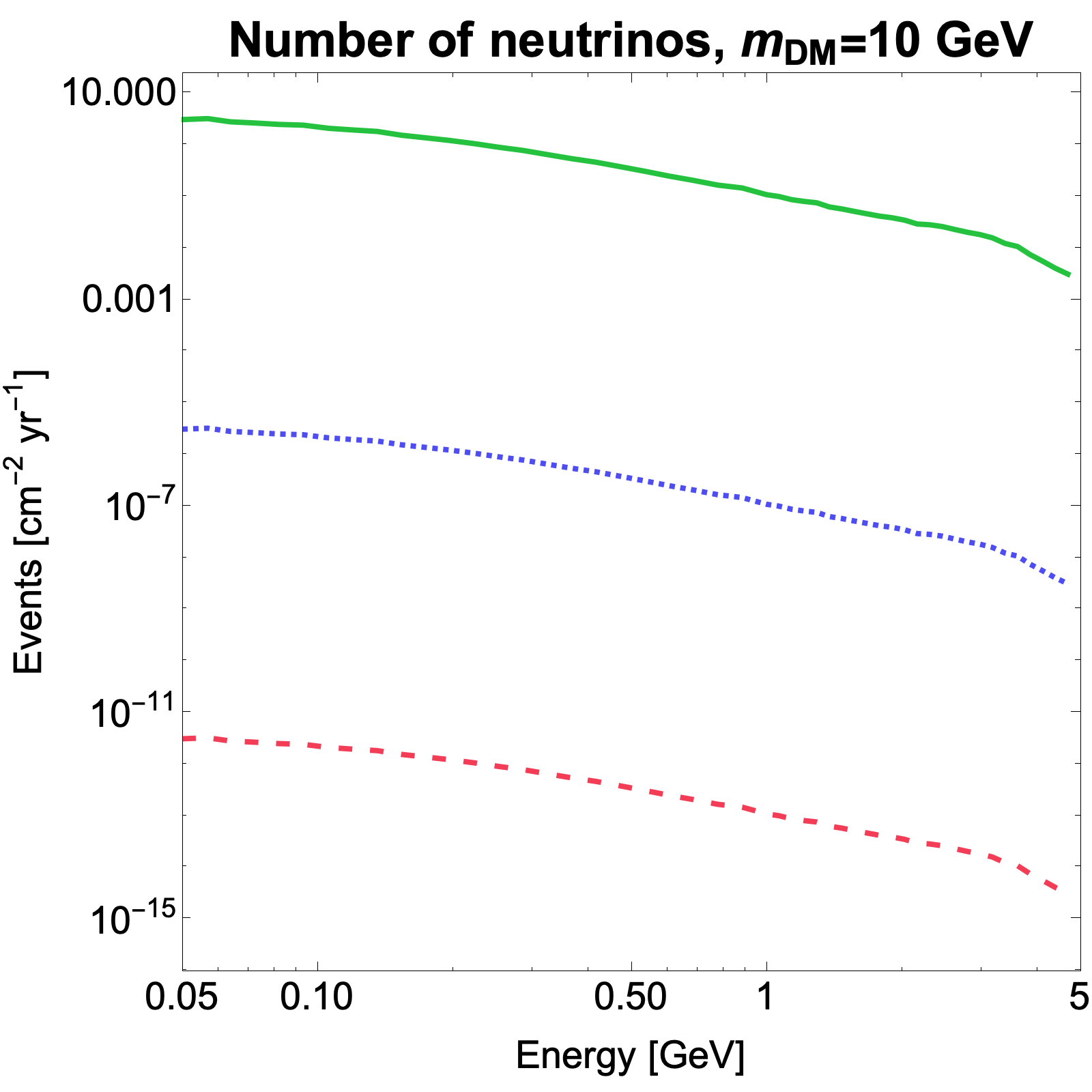}
        \subcaption{}
    \end{minipage}
    \begin{minipage}[b]{0.305\textwidth}
        \centering
        \includegraphics[width=\textwidth]{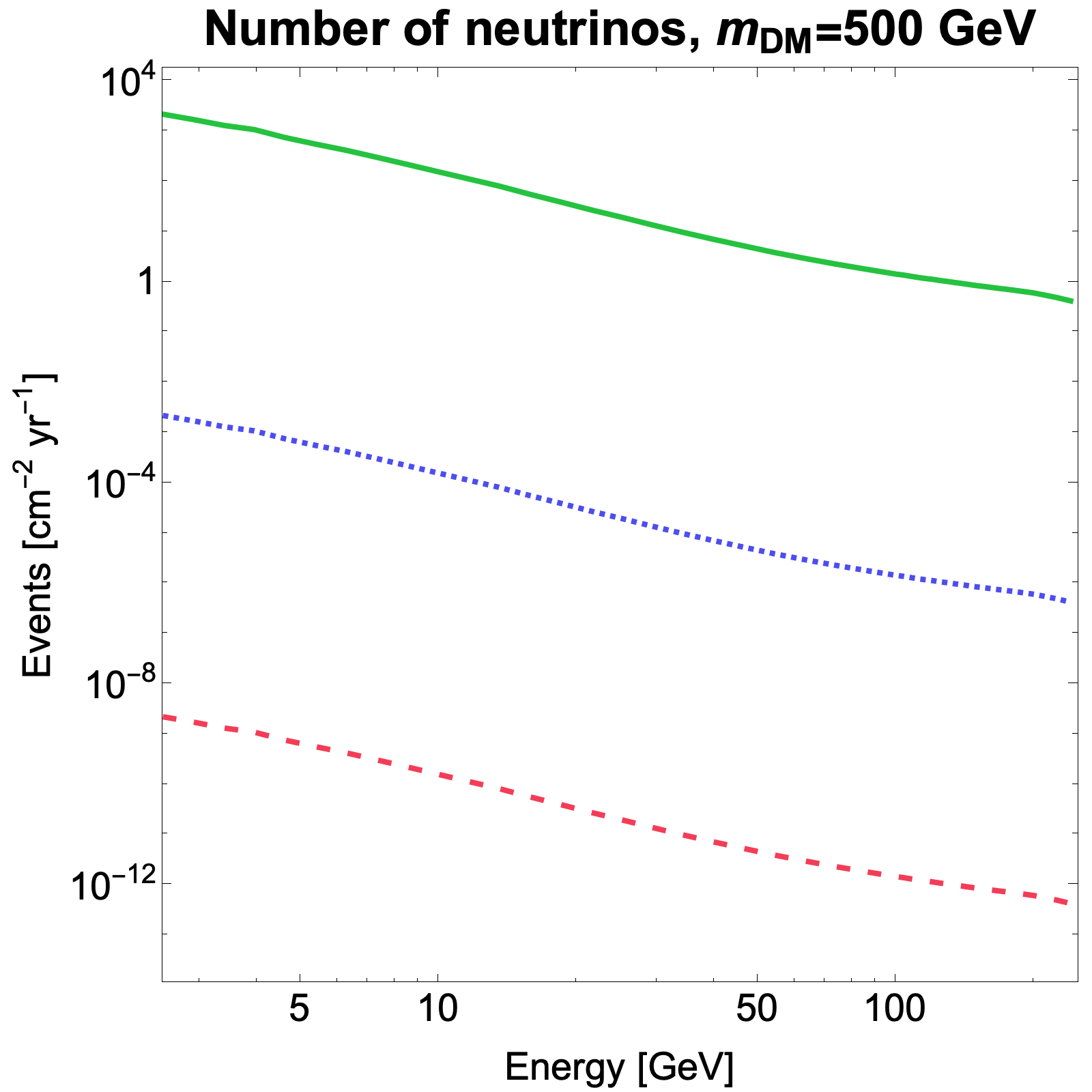}
        \subcaption{}
    \end{minipage}
    \begin{minipage}[b]{0.37\textwidth}
        \centering
        \includegraphics[width=\textwidth]{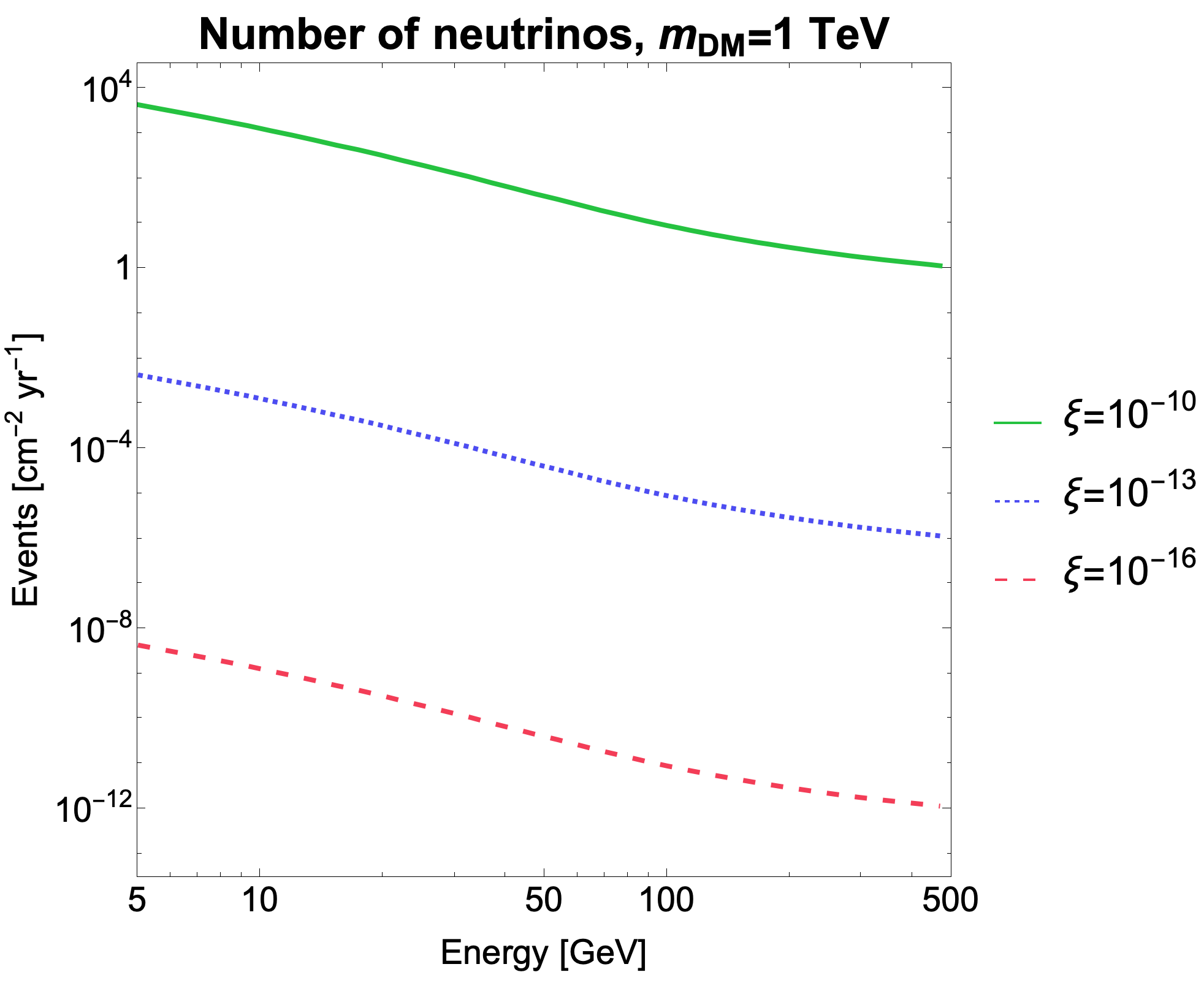}
        \subcaption{}
    \end{minipage}
    \caption{Number of neutrinos varying the mass and the coupling parameter. In these plots, the solid green line corresponds to number of muon neutrinos that can be detected on Earth, considering $\xi=10^{-10}$, the solid blue line is for $\xi=10^{-13}$ and the dotted red line corresponds to $\xi=10{-16}$. The mass for the candidate is fixed in each plot, in figure a) we consider a mass of 10 GeV, in b) 500 GeV and for c) is 1 TeV.}
\end{figure}

Again, the best case is the one with the DM mass of 1 TeV and a coupling parameter of $\xi=10^{-10}$ where the number of events is around four thousand for neutrinos with energies between 5 and 10 GeV. Similarly, for a DM mass of 500 GeV, the number of neutrinos that can be detected is close to one thousand with energies between 2.5 and 5 GeV. For the case of the mass of 10 GeV more than a year of data is needed to observe just one neutrino. 

\subsection{dSph Galaxies}
\label{sub:dsph}

These galaxies are small but, due to their characteristics, are difficult to identify against the cosmic background. They were created during the early stages of the Universe and most of the stars that comprise them are red giants ~\cite{Strigari}. 

The indirect search for dark matter in this type of object has been increasing with collaborations such as Fermi-LAT ~\cite{McDaniel:2023bju}. The Dark Energy Survey (DES) ~\cite{DES:2018lct} explores the universe using certain wavelengths to study its expansion and has found that dSphs contain large amounts of dark matter and lack gas. Recently, experiments seeking signals from dark matter have mainly used gamma rays with energy ranges from 100 MeV to 100 TeV. In this range, DM should be a thermal relic that is consistent with the measured value. 

\subsubsection{Gamma Rays}

Module \texttt{h2} was used to compute the differential flux produced in the dSphs shown in the table ~\ref{tab:infor}. We followed the same procedure as for the Milky Way to obtain the differential flux; that is, the same module for different values in the range of the mass of the candidate was run for these objects. 

\begin{figure}
    \centering
    \begin{minipage}[b]{0.295\textwidth}
        \centering
        \includegraphics[width=\textwidth]{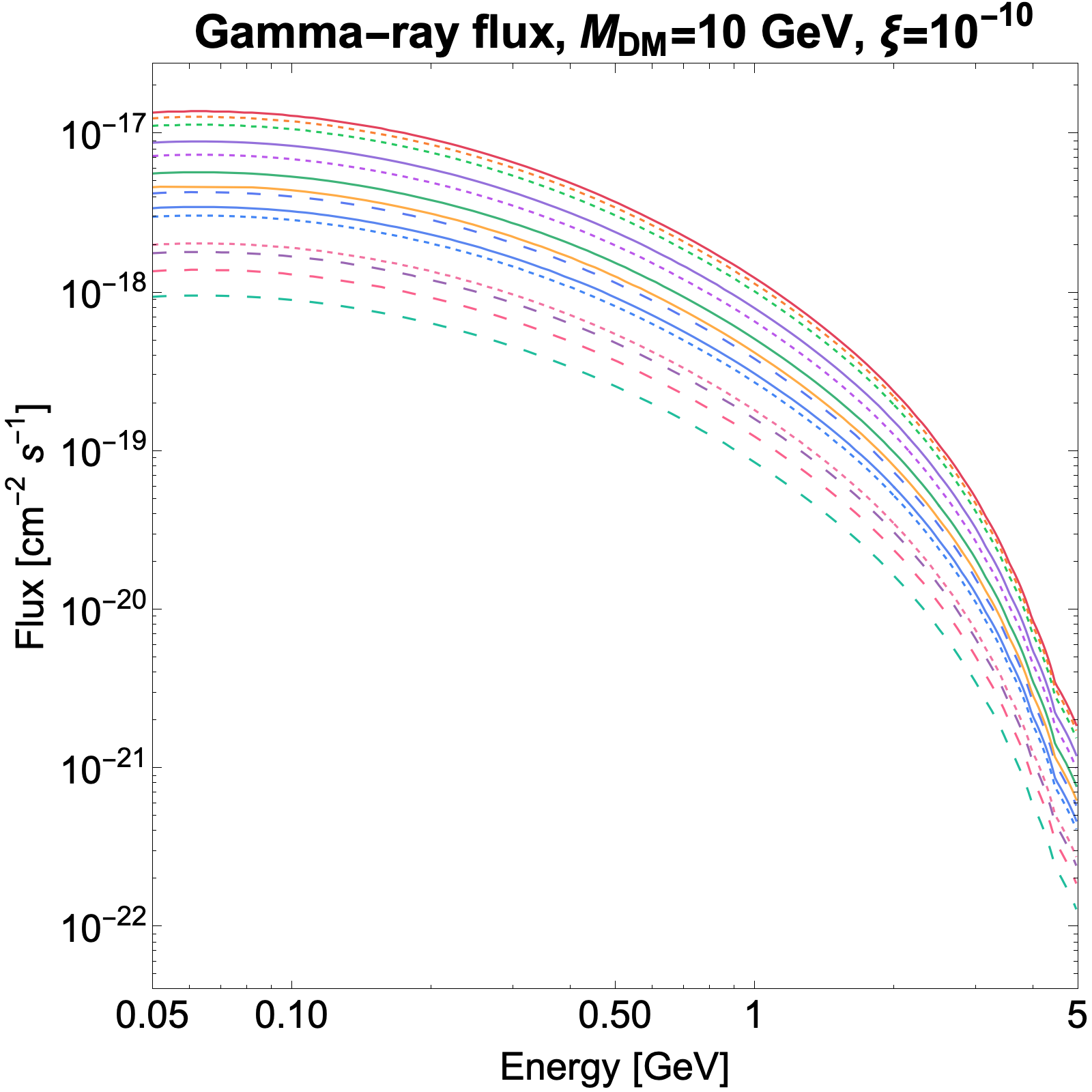}
        \subcaption{}
    \end{minipage}
    \begin{minipage}[b]{0.295\textwidth}
        \centering
        \includegraphics[width=\textwidth]{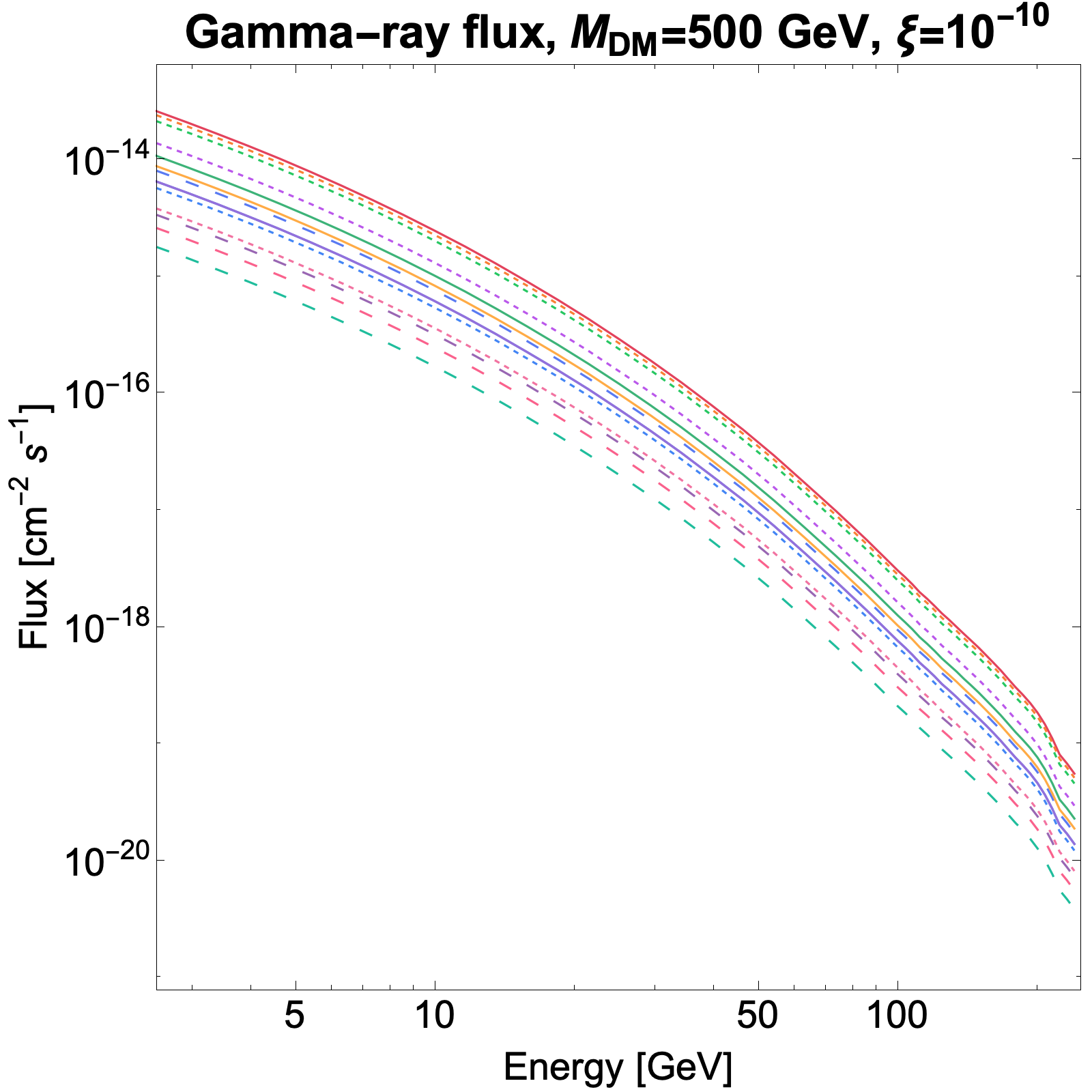}
        \subcaption{}
    \end{minipage}
    \begin{minipage}[b]{0.39\textwidth}
        \centering
        \includegraphics[width=\textwidth]{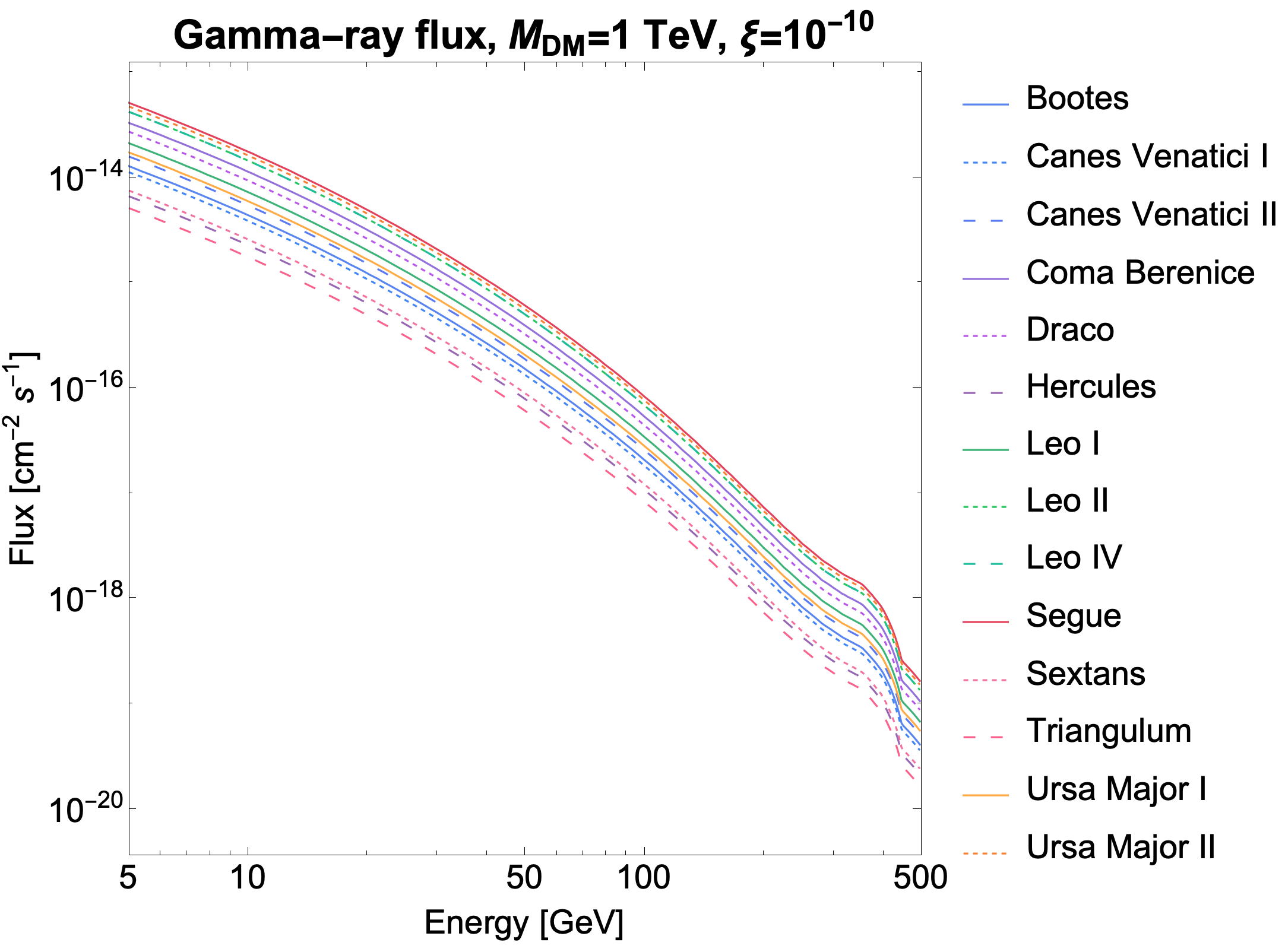}
        \subcaption{}
    \end{minipage}

    \vspace{0.5em}

    \begin{minipage}[b]{0.295\textwidth}
        \centering
        \includegraphics[width=\textwidth]{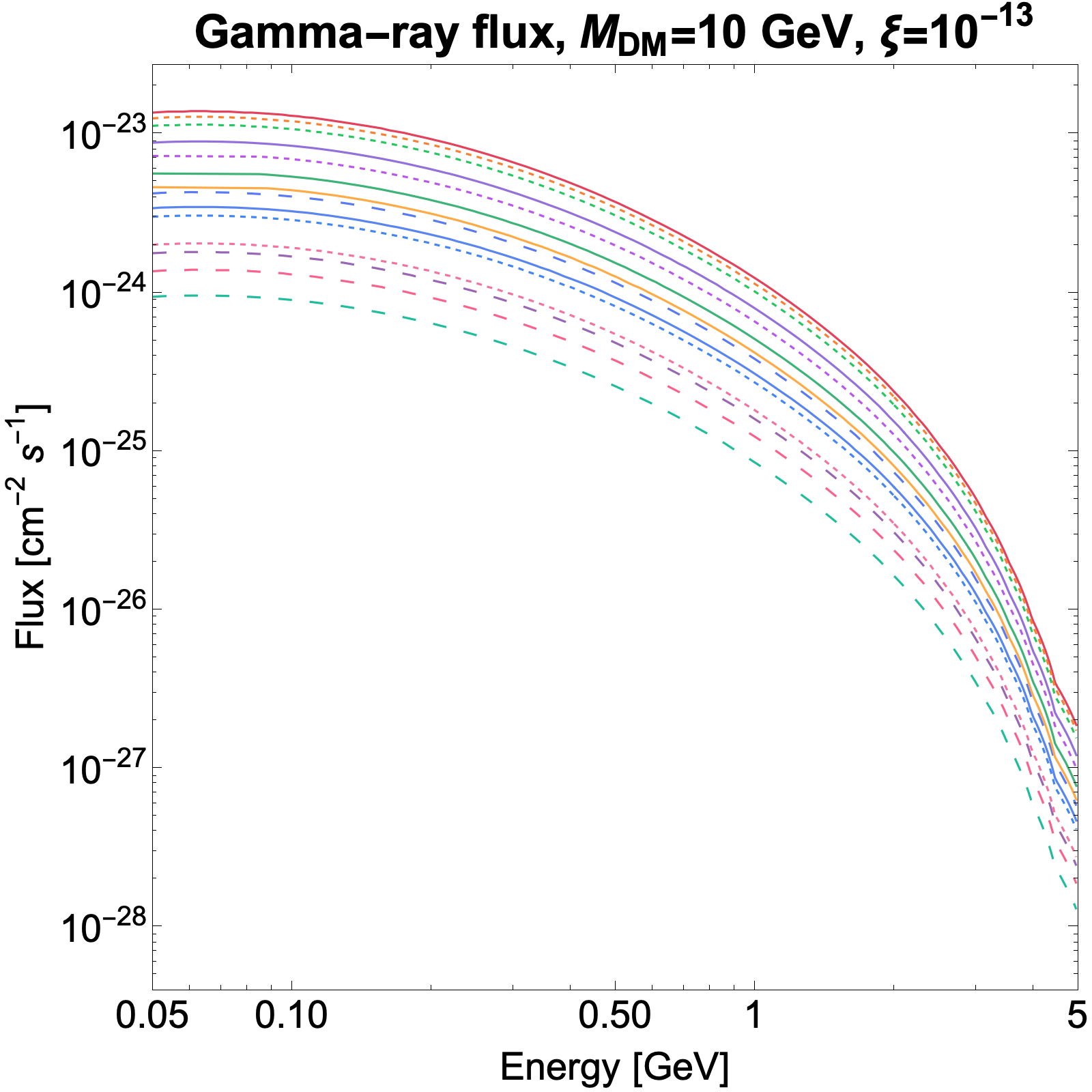}
        \subcaption{}
    \end{minipage}
    \begin{minipage}[b]{0.295\textwidth}
        \centering
        \includegraphics[width=\textwidth]{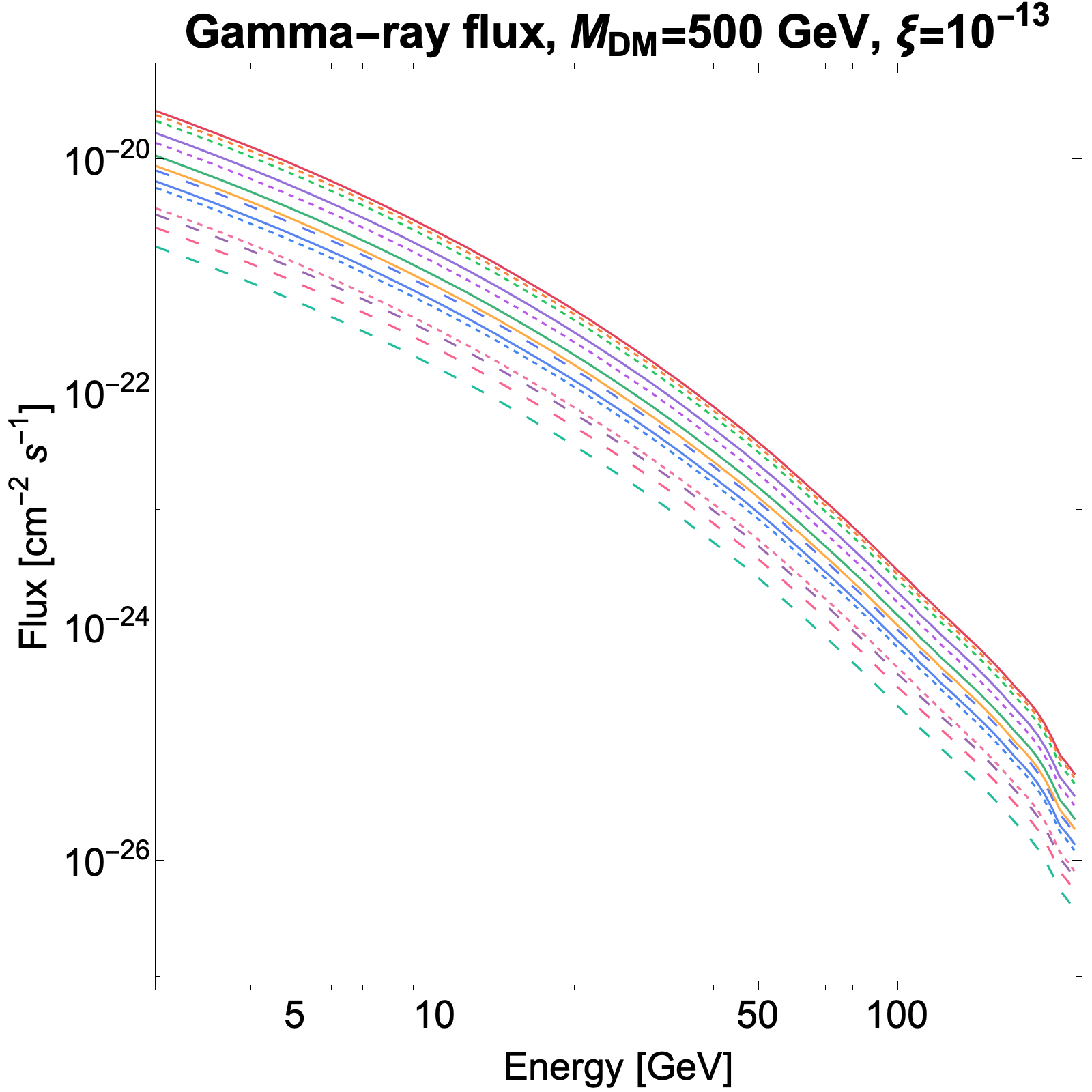}
        \subcaption{}
    \end{minipage}
    \begin{minipage}[b]{0.39\textwidth}
        \centering
        \includegraphics[width=\textwidth]{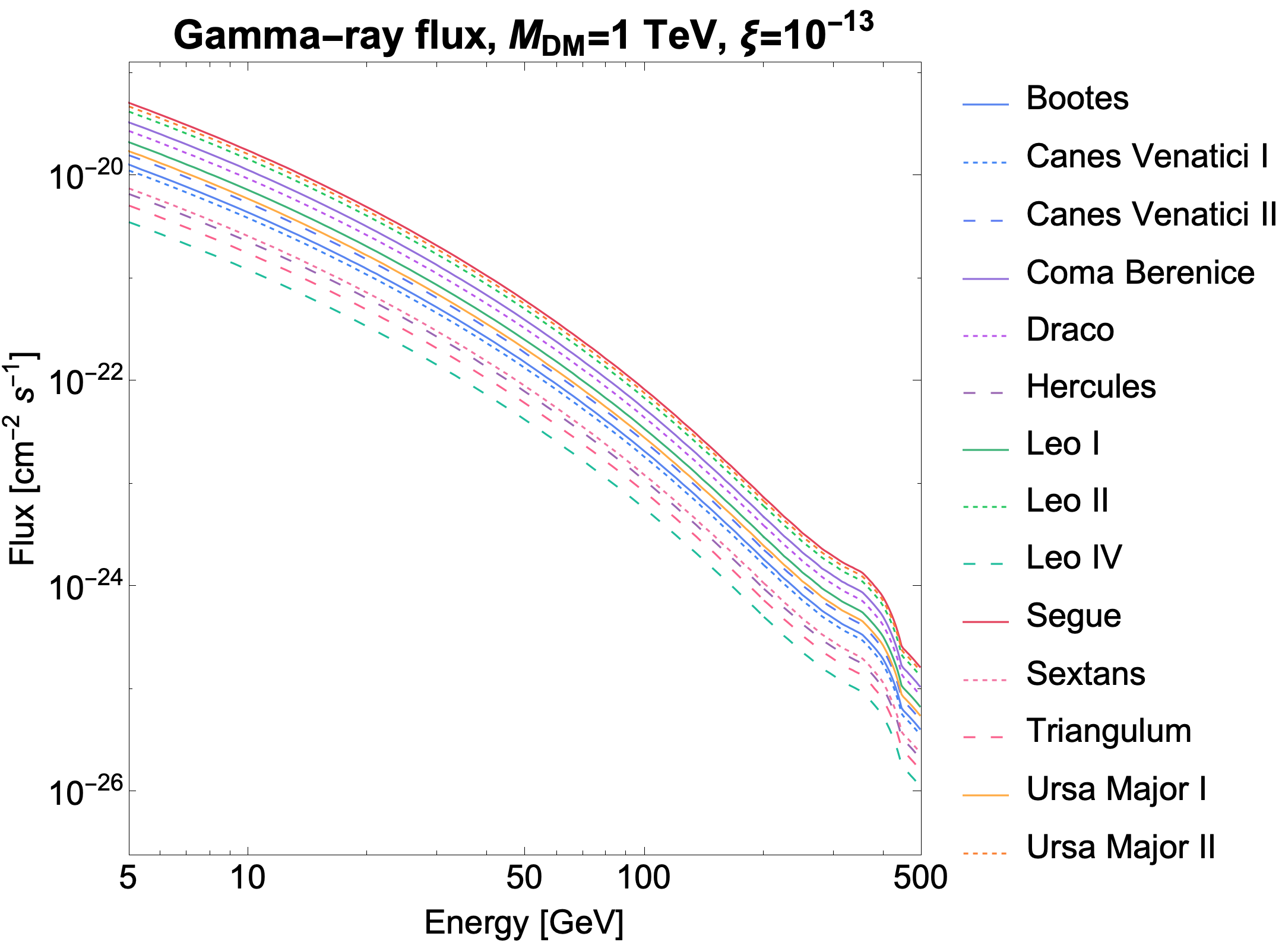}
        \subcaption{}
    \end{minipage}

    \vspace{0.5em}

    \begin{minipage}[b]{0.295\textwidth}
        \centering
        \includegraphics[width=\textwidth]{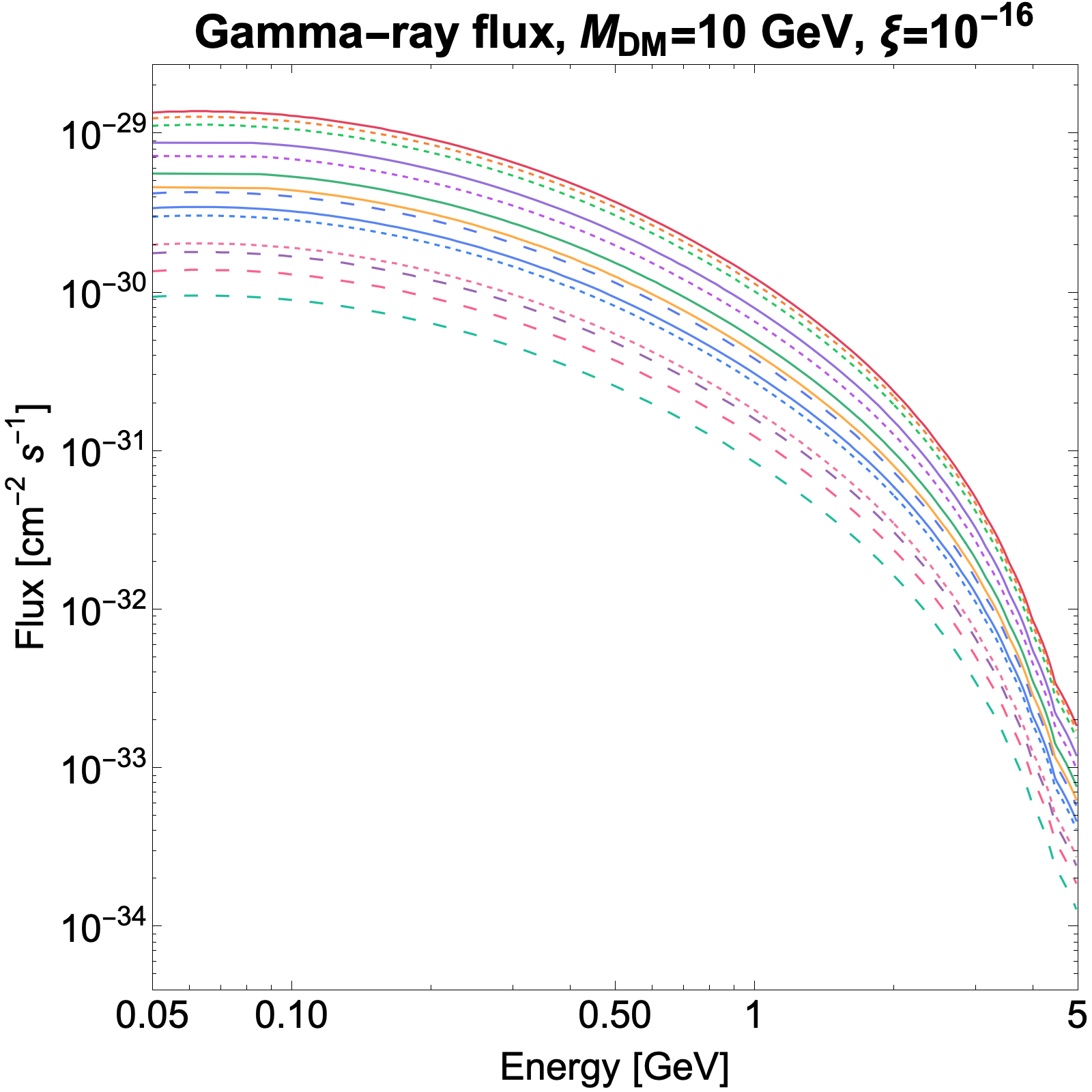}
        \subcaption{}
    \end{minipage}
    \begin{minipage}[b]{0.295\textwidth}
        \centering
        \includegraphics[width=\textwidth]{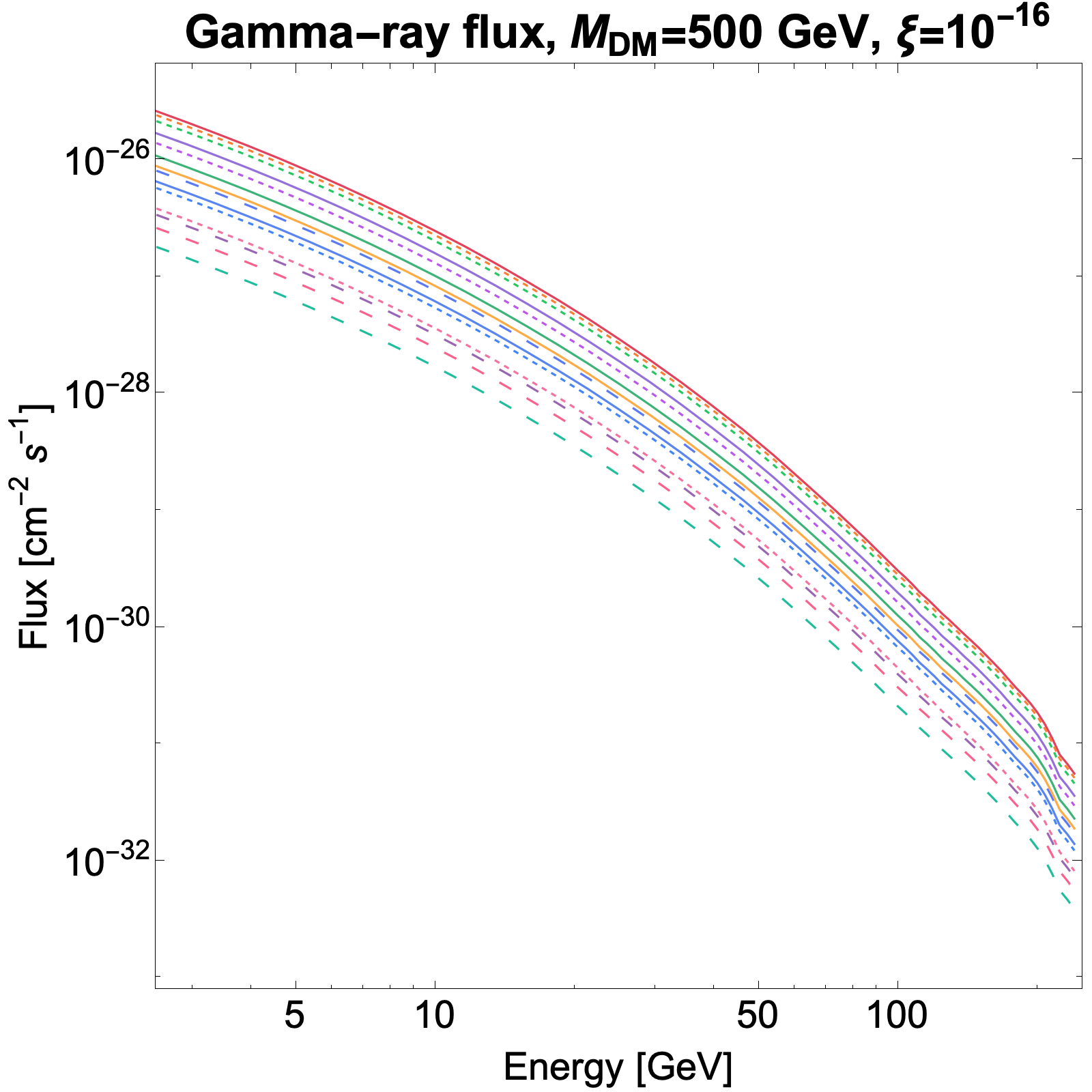}
        \subcaption{}
    \end{minipage}
    \begin{minipage}[b]{0.39\textwidth}
        \centering
        \includegraphics[width=\textwidth]{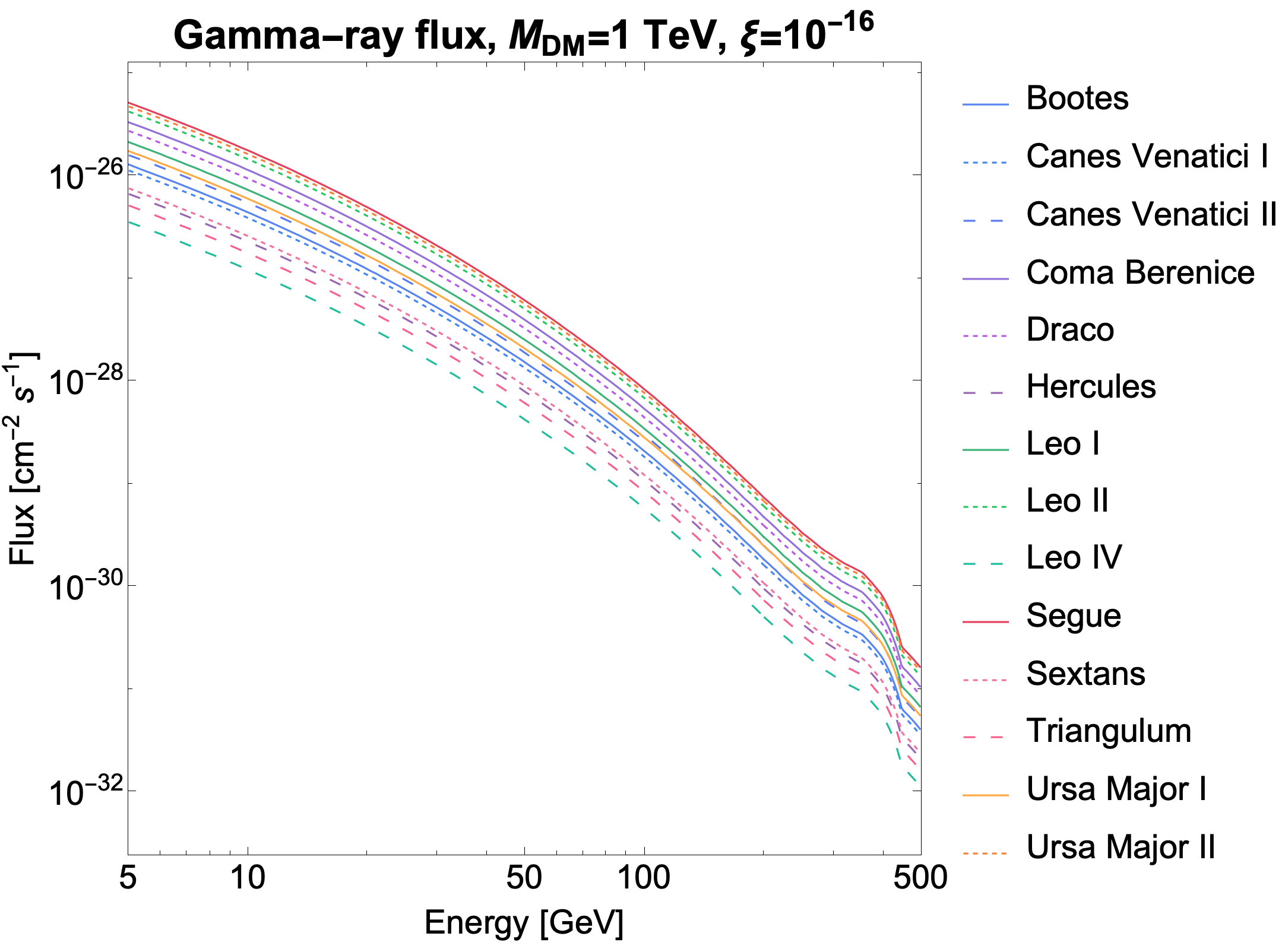}
        \subcaption{}
    \end{minipage}
    \caption{Flux of gamma rays produced in some dSphs. Each line corresponds to one dSph. For each plot there is a fixed value for the mass of the candidate and the coupling parameter. In the first row $\xi=10^{-10}$, in the second is $10^{-13}$ and for the third one is $10^{-16}$. And for each column the mass is 10 GeV in first one, 100 GeV in the second one and 1 TeV in the last one.}
    \label{fig:Ga-flux-dSph}
    \vspace{0.5em}
\end{figure}

For the gamma rays, the behavior is similar to those produced in our galaxy, as the flux decreases as the mass value increases. The relation between the plots in the first and second columns is approximately $2.4\times10^5$, and between the second and third columns is only 24, this relation is the same for all the values of $\xi$. 

\begin{figure}
    \centering
    \begin{minipage}[b]{0.295\textwidth}
        \centering
        \includegraphics[width=\textwidth]{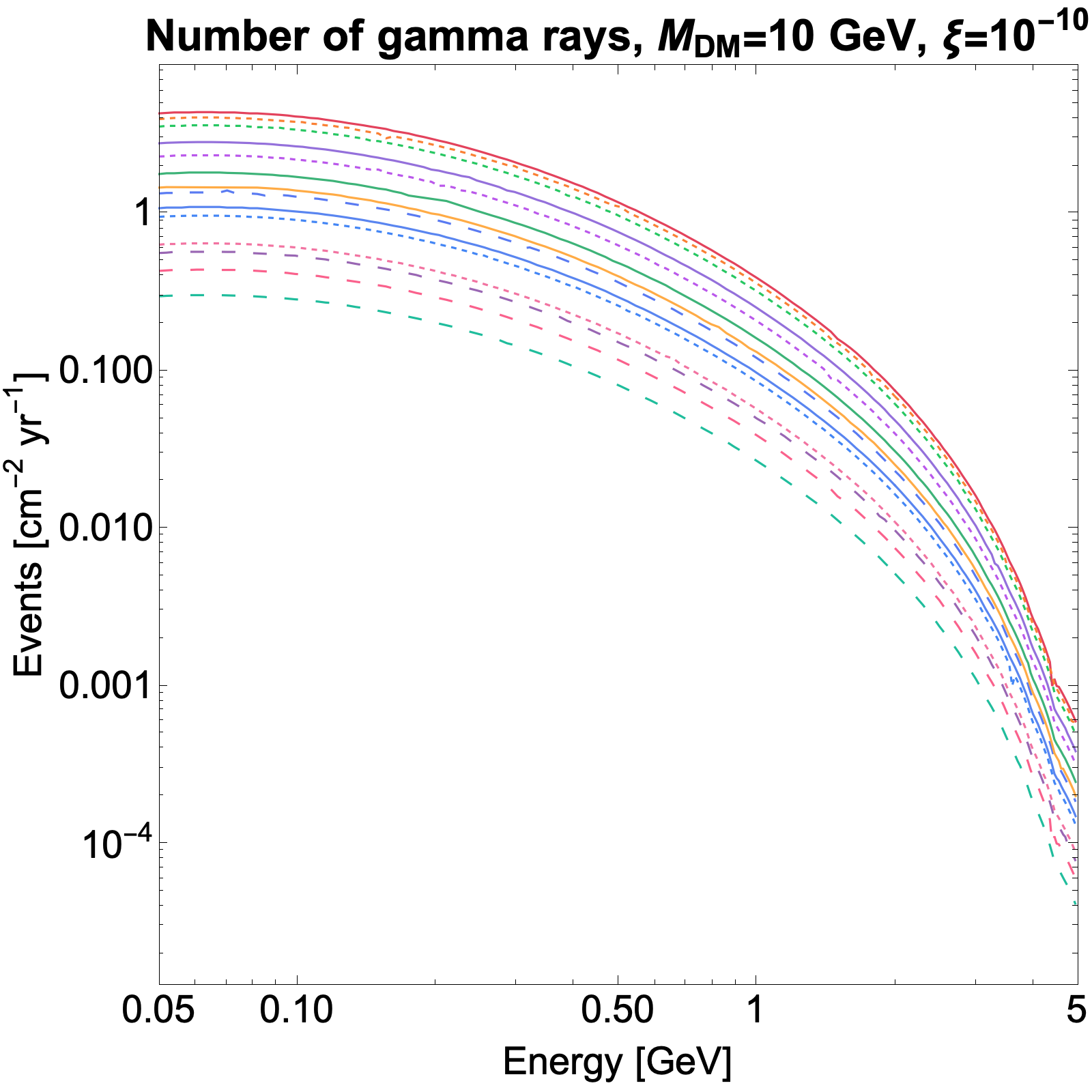}
        \subcaption{}
    \end{minipage}
    \begin{minipage}[b]{0.295\textwidth}
        \centering
        \includegraphics[width=\textwidth]{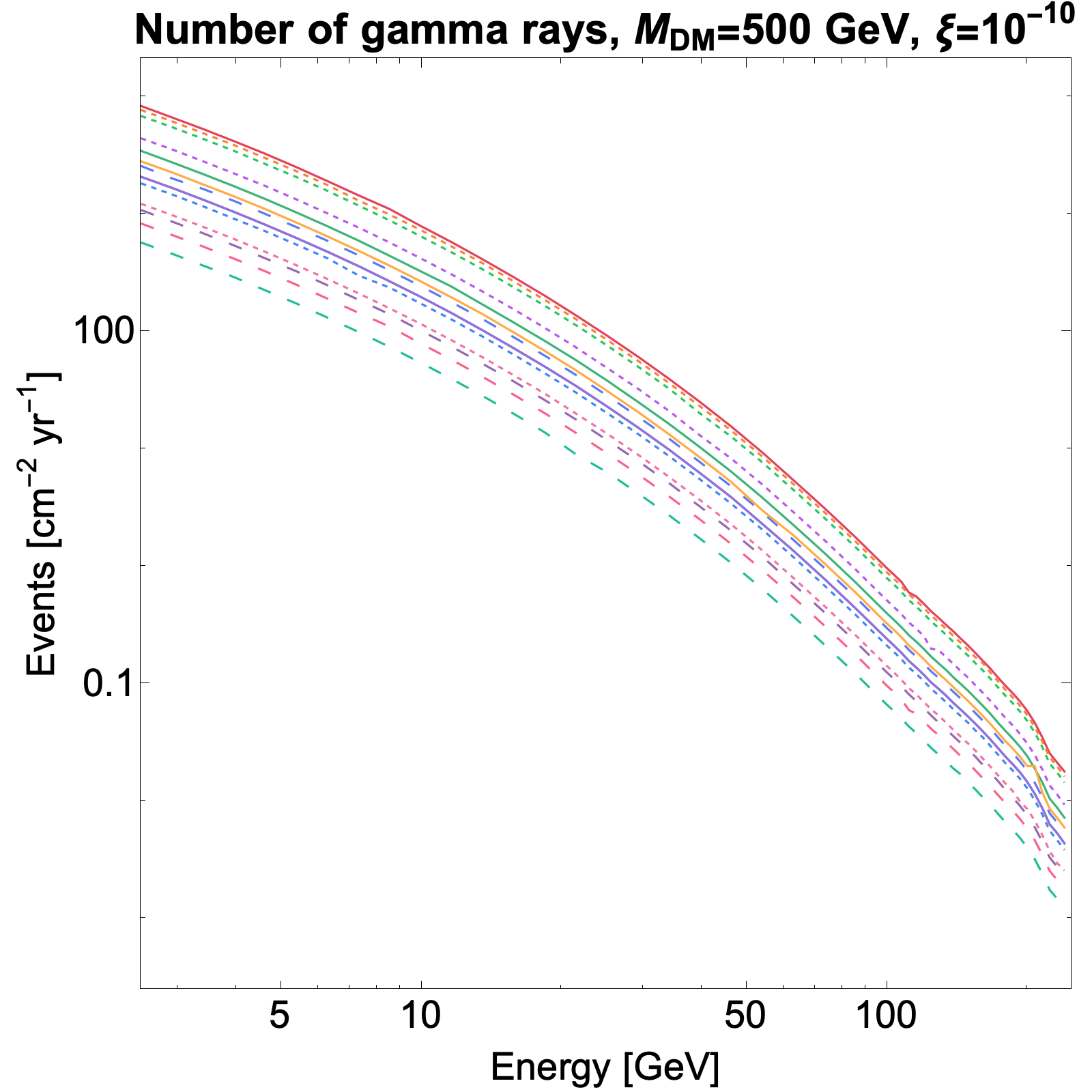}
        \subcaption{}
    \end{minipage}
    \begin{minipage}[b]{0.39\textwidth}
        \centering
        \includegraphics[width=\textwidth]{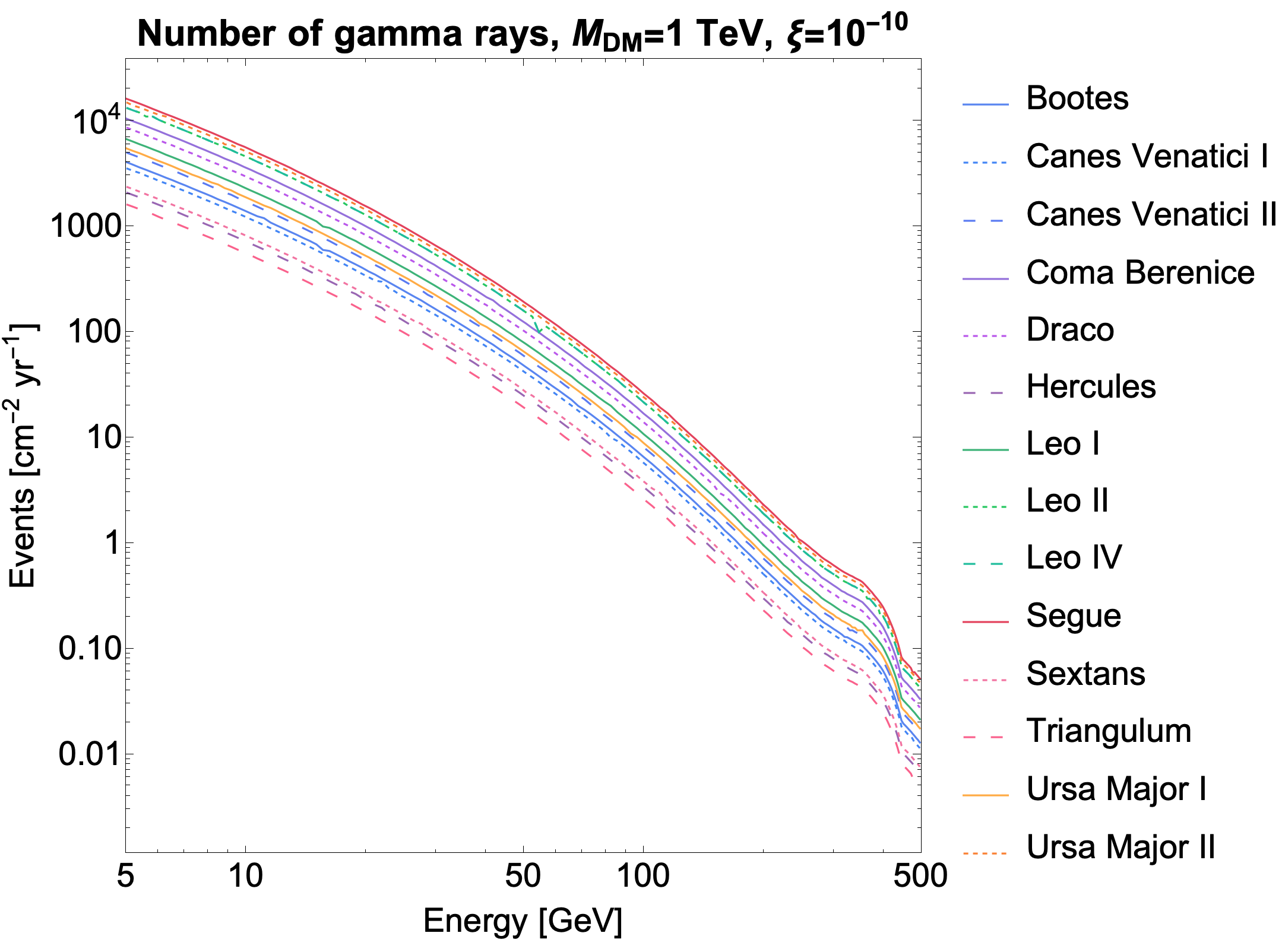}
        \subcaption{}
    \end{minipage}

    \vspace{0.5em}

    \begin{minipage}[b]{0.295\textwidth}
        \centering
        \includegraphics[width=\textwidth]{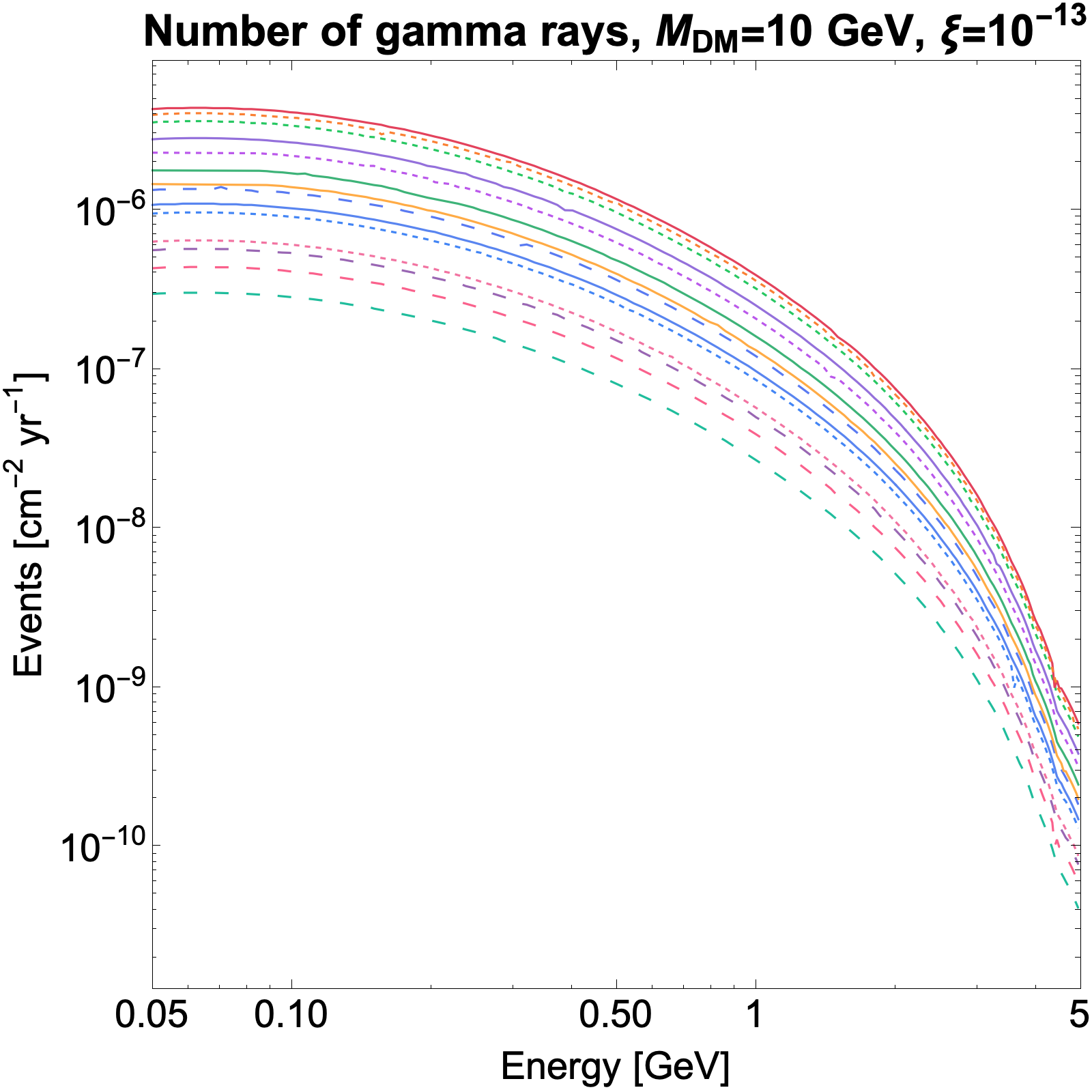}
        \subcaption{}
    \end{minipage}
    \begin{minipage}[b]{0.295\textwidth}
        \centering
        \includegraphics[width=\textwidth]{Images/dSph/Ga_Eventos_xi10_M500.png}
        \subcaption{}
    \end{minipage}
    \begin{minipage}[b]{0.39\textwidth}
        \centering
        \includegraphics[width=\textwidth]{Images/dSph/Ga_Eventos_xi10_M1000.png}
        \subcaption{}
    \end{minipage}

    \vspace{0.5em}

    \begin{minipage}[b]{0.295\textwidth}
        \centering
        \includegraphics[width=\textwidth]{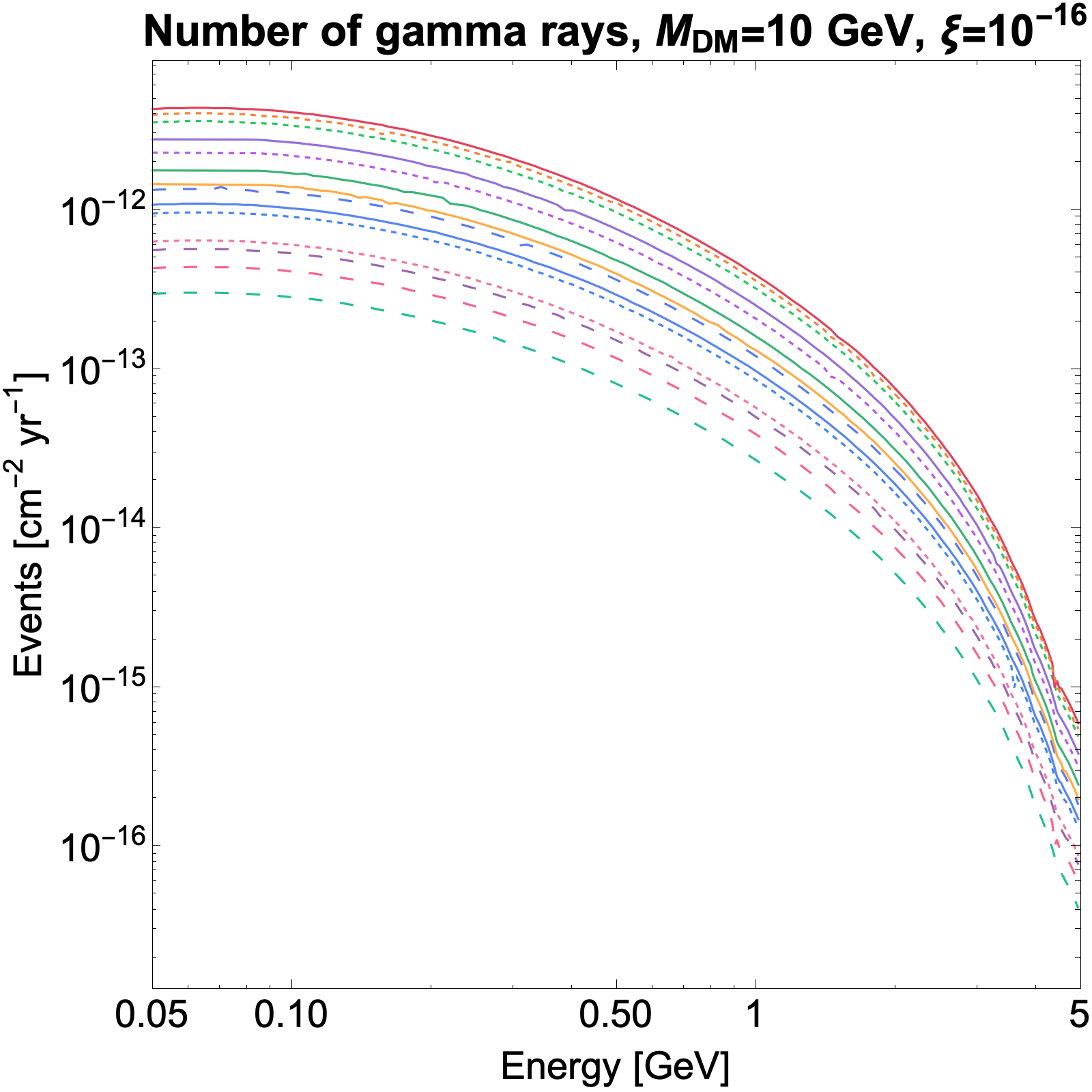}
        \subcaption{}
    \end{minipage}
    \begin{minipage}[b]{0.295\textwidth}
        \centering
        \includegraphics[width=\textwidth]{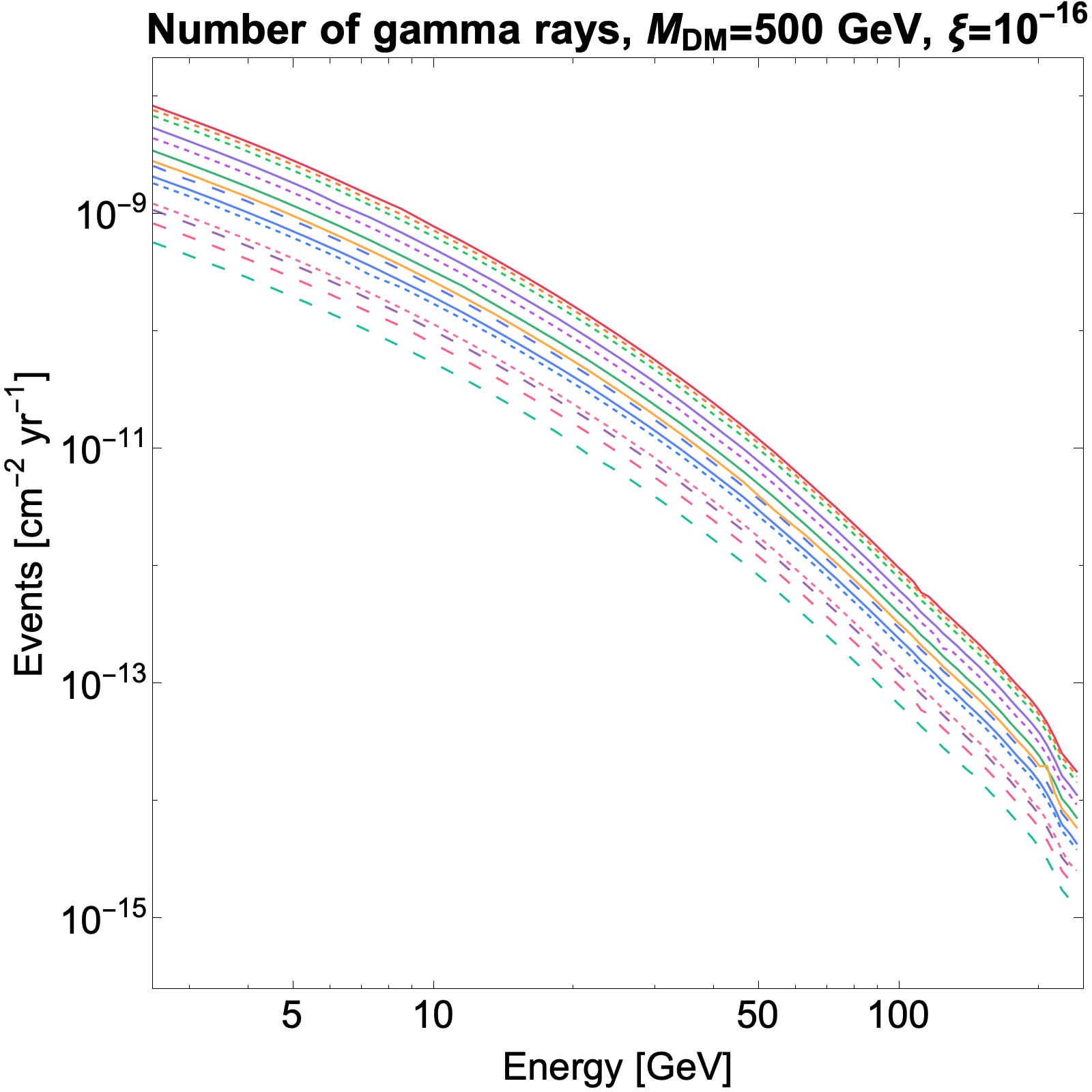}
        \subcaption{}
    \end{minipage}
    \begin{minipage}[b]{0.39\textwidth}
        \centering
        \includegraphics[width=\textwidth]{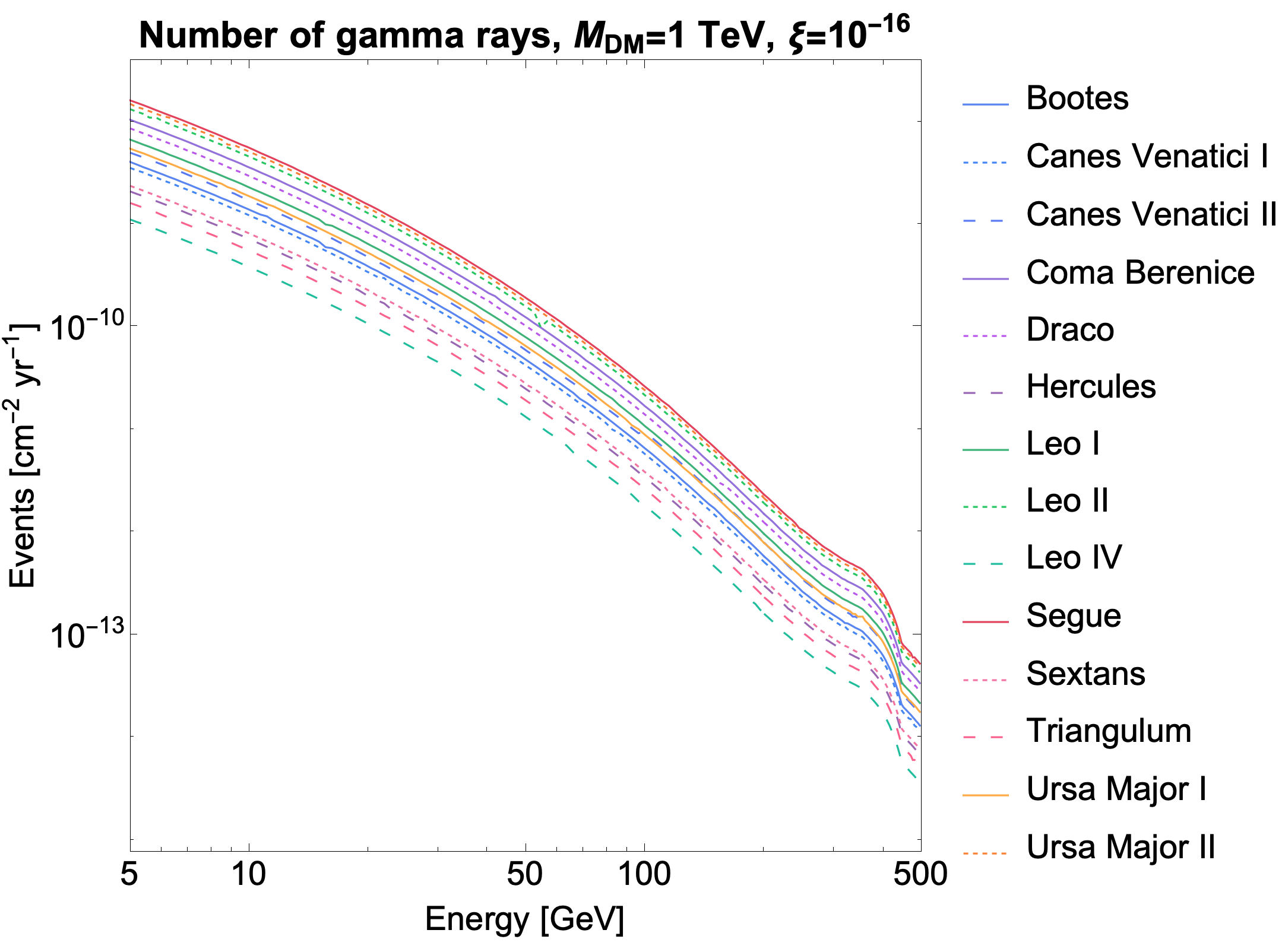}
        \subcaption{}
    \end{minipage}
    \caption{Number of gamma rays produced in the dSphs selected for this analysis. Each line corresponds to one dSph. For each plot, there is a fixed value for the mass of the candidate and the coupling parameter. In the first row $\xi=10^{-10}$, in the second is $10^{-13}$ and for the third one is $10^{-16}$. And for each column the mass is 10 GeV in first one, 100 GeV in the second one and 1 TeV in the last one.}
    \label{fig:Ga-eve-dSph}
    \vspace{0.5em}
\end{figure}

We also estimated the number of gamma rays with a specific energy considering a detector with one squared kilometer of effective area and one year data. The plots obtained are shown in Figure ~\ref{fig:Ga-eve-dSph}.

Again, the best scenario is the combination of $m=10$ GeV and coupling parameter of $10^{-10}$, where the possible number is between one to ten thousand events with energy between 5 and 10 GeV. For the lowest value of the coupling parameter, $\xi=10^{-16}$ the time required to detect one gamma ray is of various years. 

\subsubsection{Neutrinos}

\begin{figure}
    \centering
    \begin{minipage}[b]{0.295\textwidth}
        \centering
        \includegraphics[width=\textwidth]{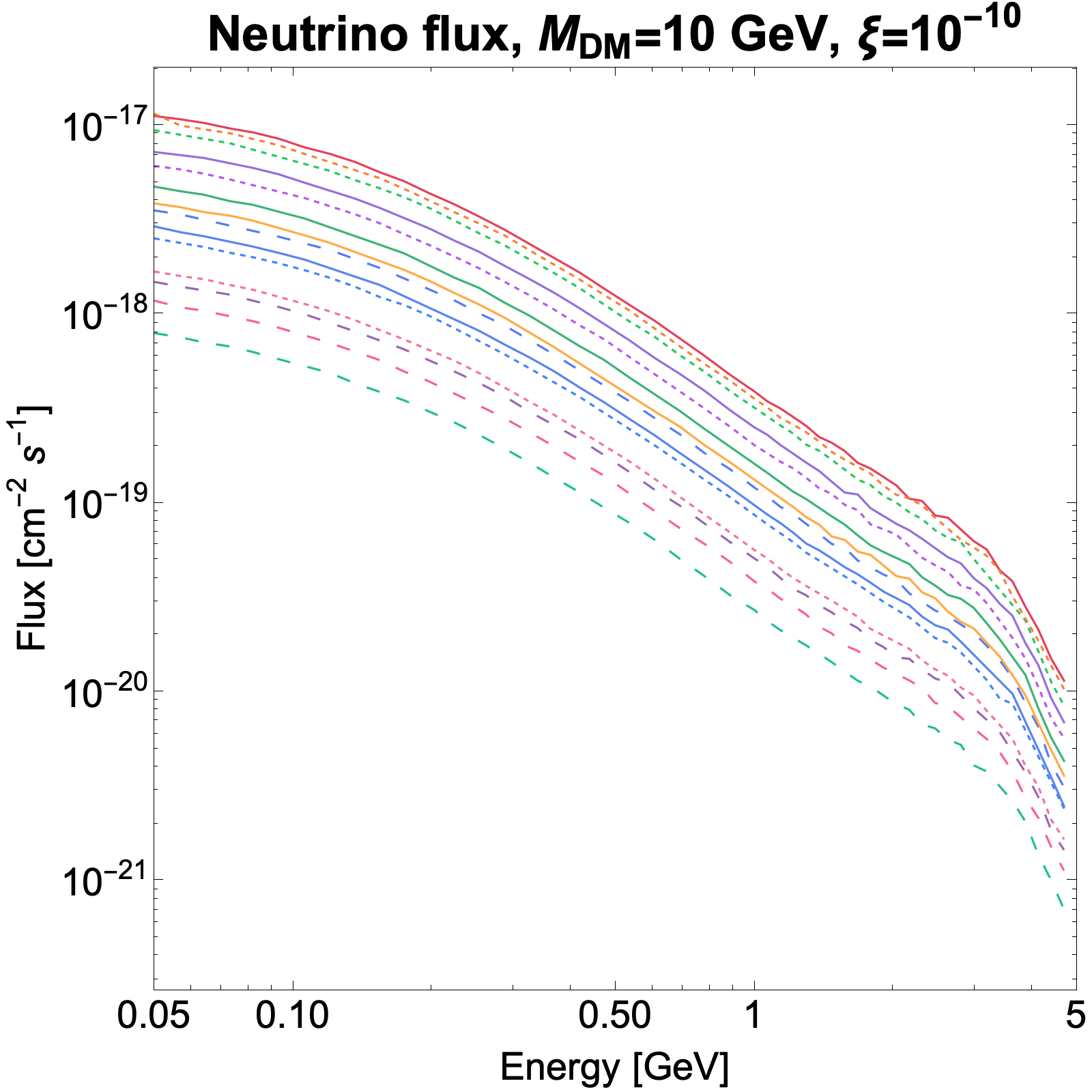}
        \subcaption{}
    \end{minipage}
    \begin{minipage}[b]{0.295\textwidth}
        \centering
        \includegraphics[width=\textwidth]{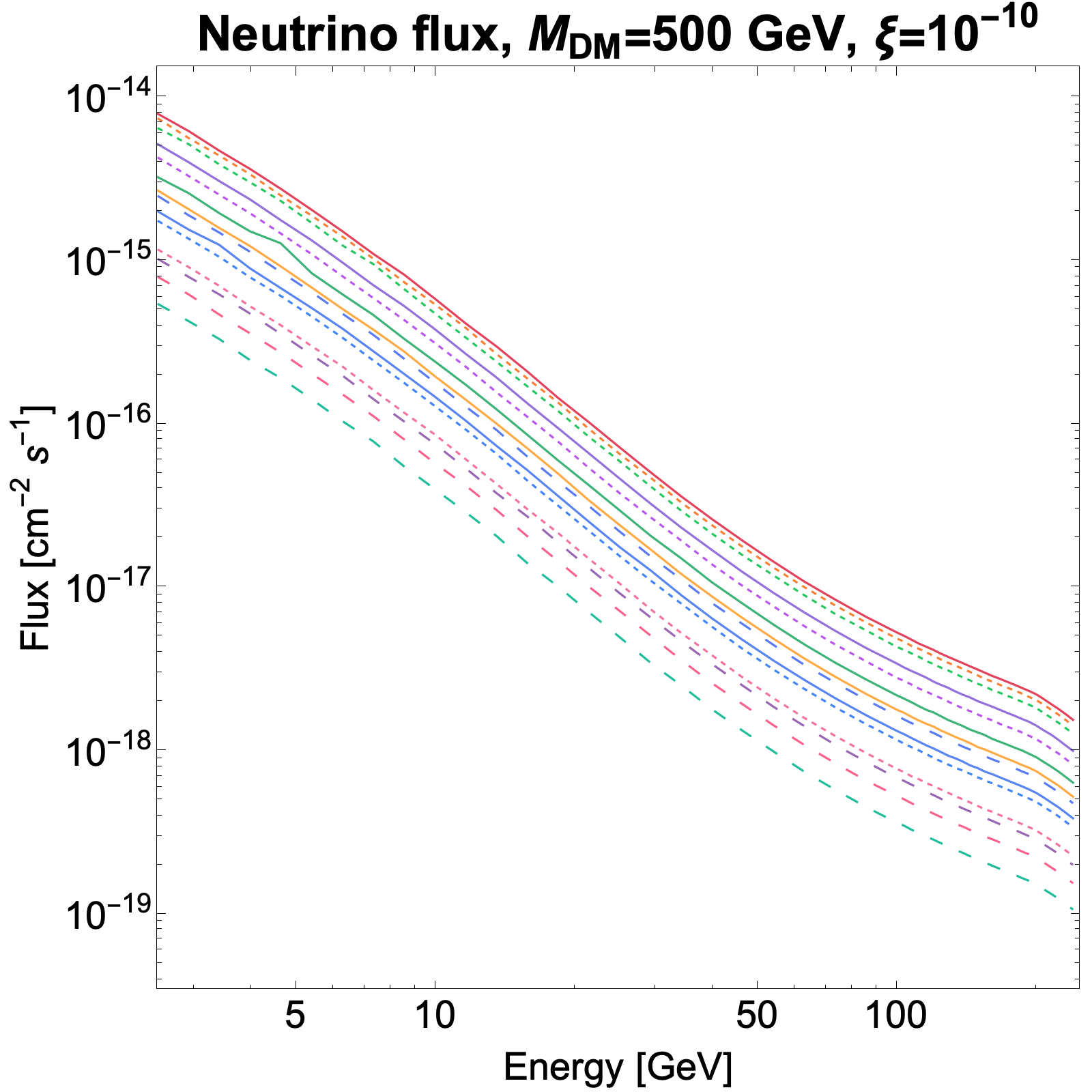}
        \subcaption{}
    \end{minipage}
    \begin{minipage}[b]{0.39\textwidth}
        \centering
        \includegraphics[width=\textwidth]{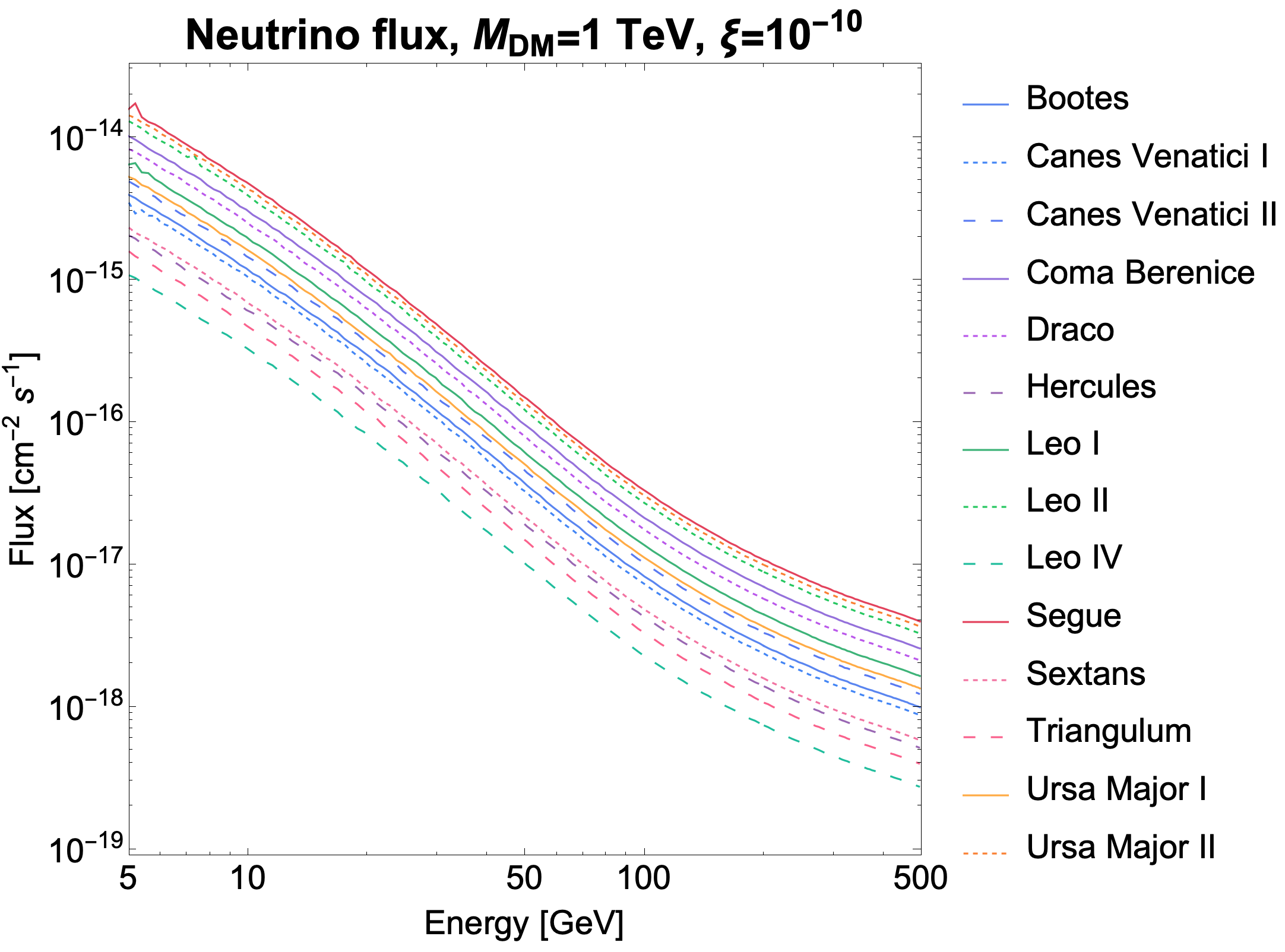}
        \subcaption{}
    \end{minipage}

    \vspace{0.5em}

    \begin{minipage}[b]{0.295\textwidth}
        \centering
        \includegraphics[width=\textwidth]{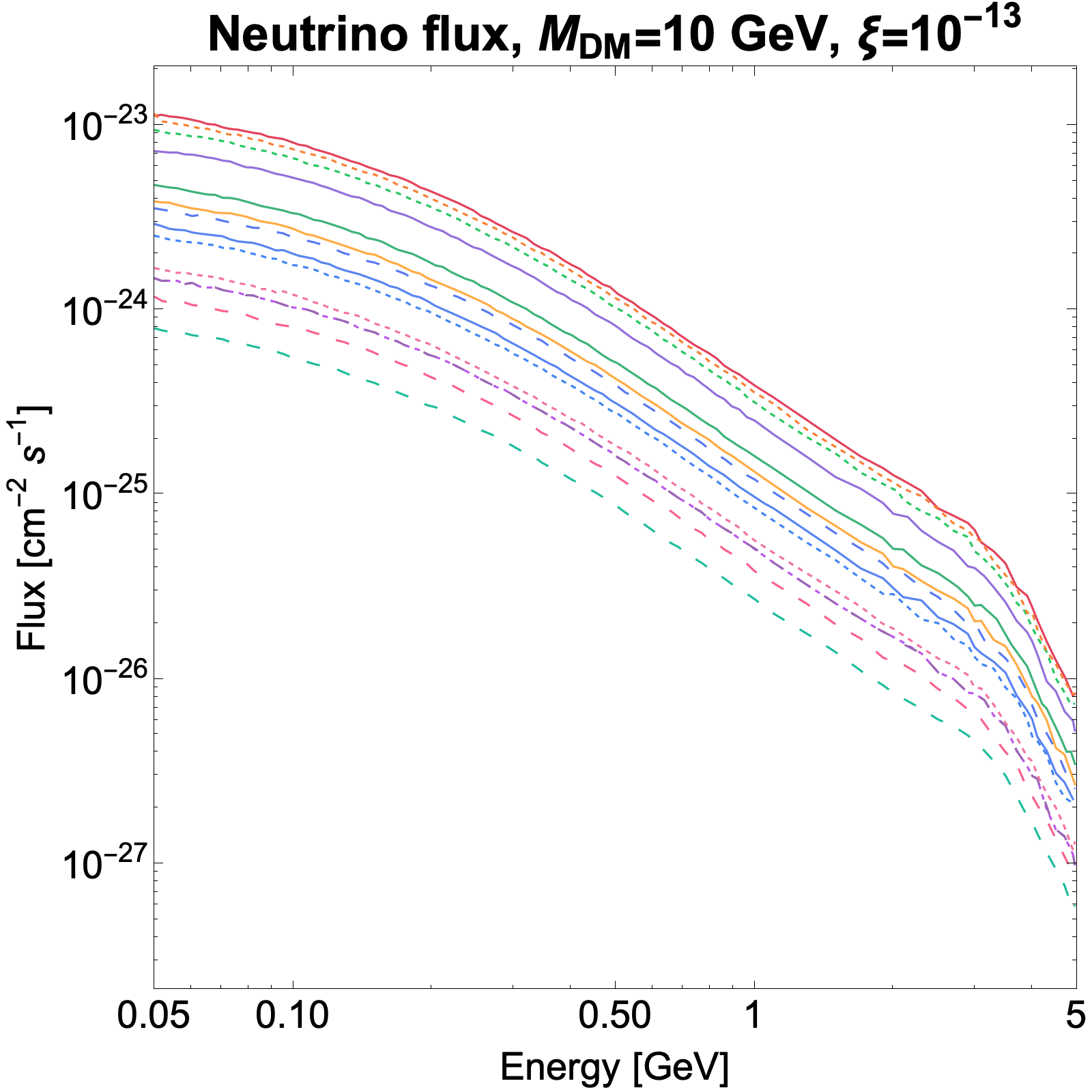}
        \subcaption{}
    \end{minipage}
    \begin{minipage}[b]{0.295\textwidth}
        \centering
        \includegraphics[width=\textwidth]{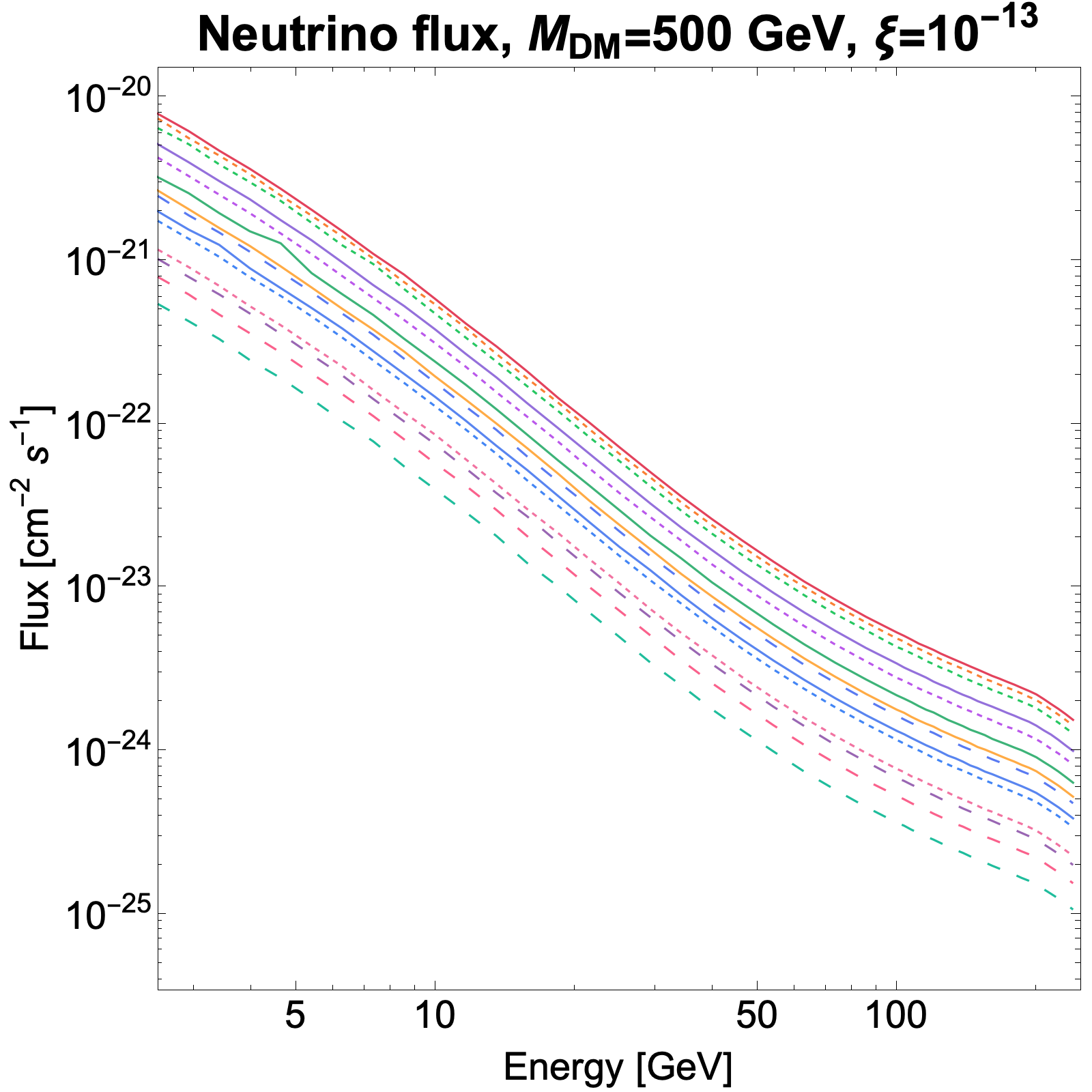}
        \subcaption{}
    \end{minipage}
    \begin{minipage}[b]{0.39\textwidth}
        \centering
        \includegraphics[width=\textwidth]{Images/dSph/Nu_Flujo_xi10_M1000.png}
        \subcaption{}
    \end{minipage}

    \vspace{0.5em}

    \begin{minipage}[b]{0.295\textwidth}
        \centering
        \includegraphics[width=\textwidth]{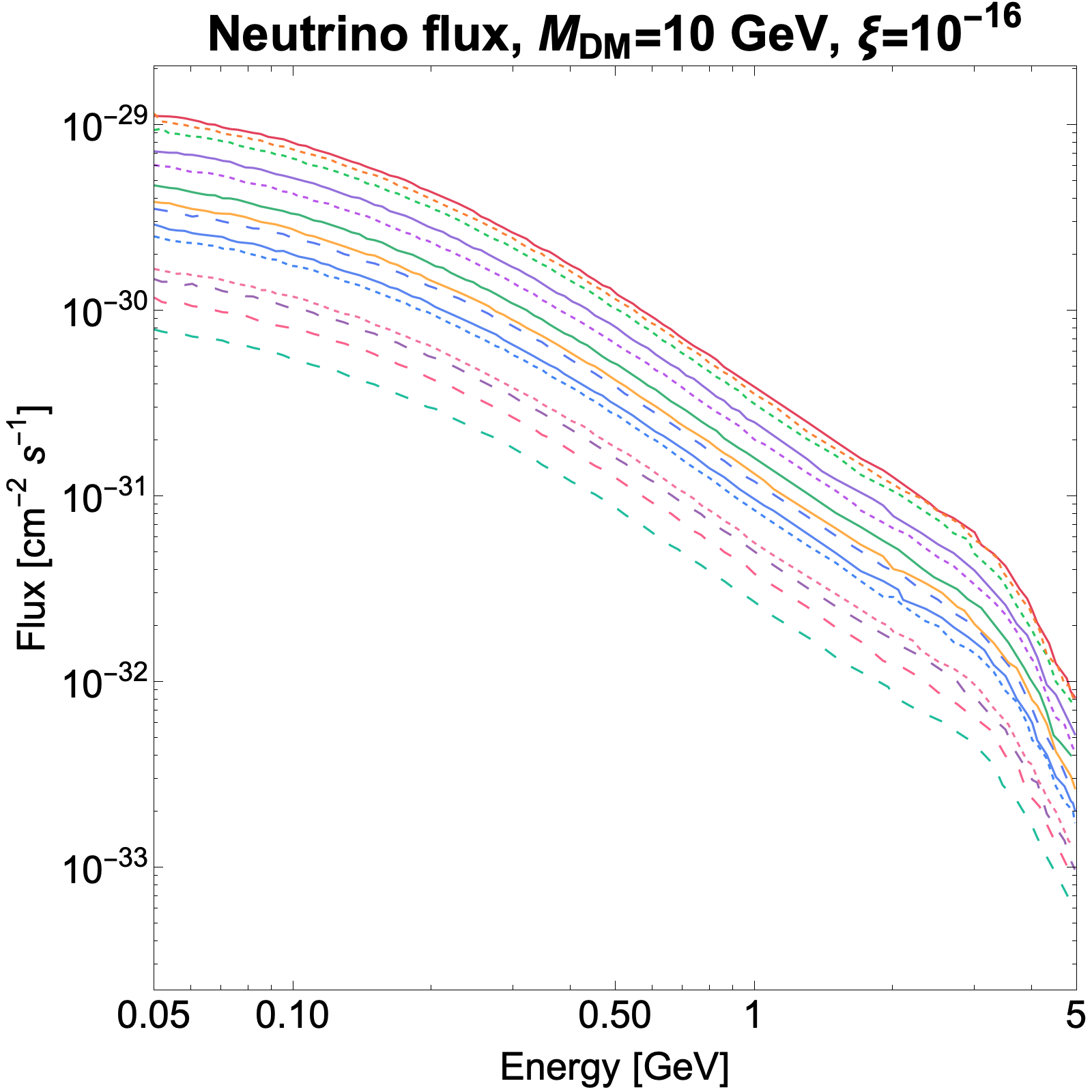}
        \subcaption{}
    \end{minipage}
    \begin{minipage}[b]{0.295\textwidth}
        \centering
        \includegraphics[width=\textwidth]{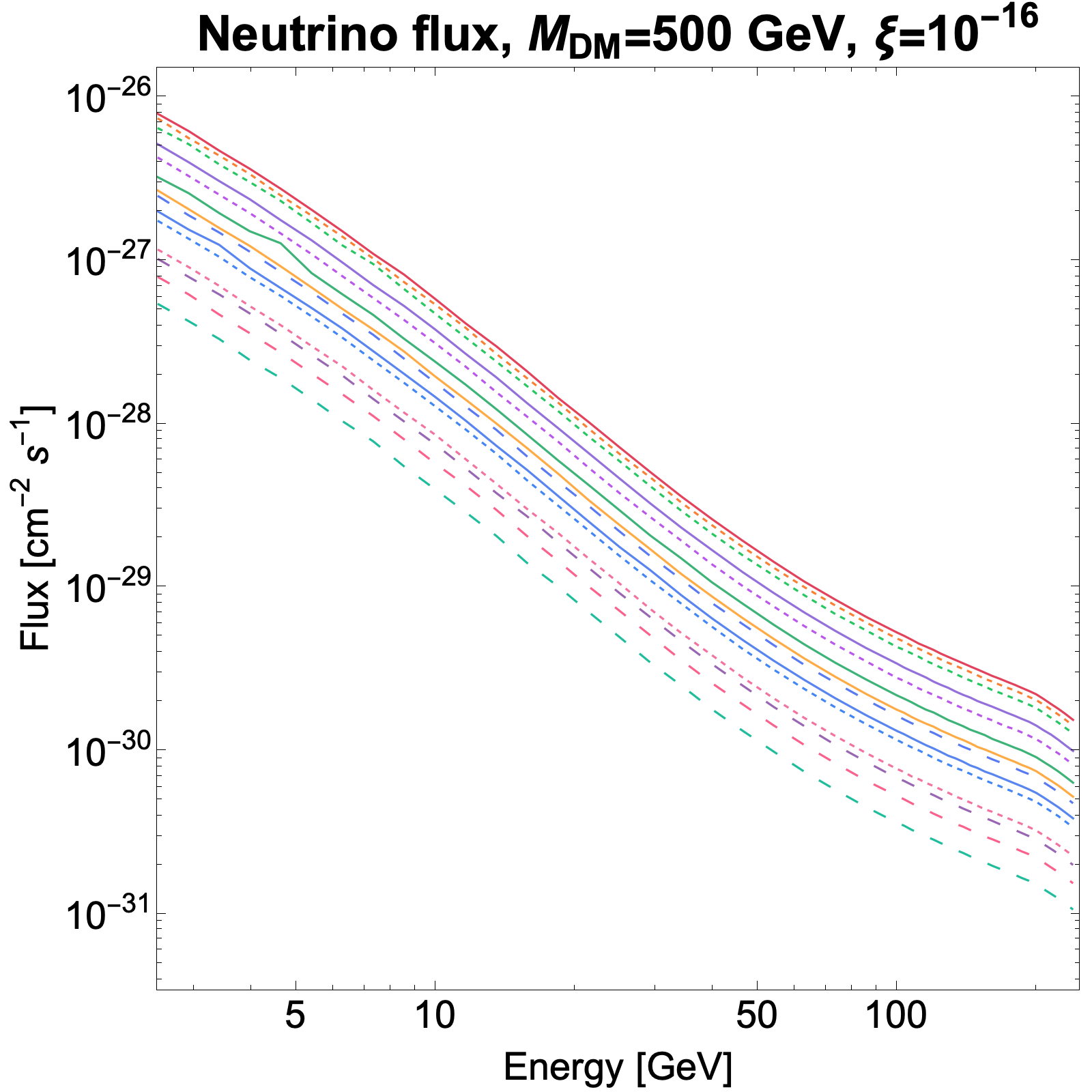}
        \subcaption{}
    \end{minipage}
    \begin{minipage}[b]{0.39\textwidth}
        \centering
        \includegraphics[width=\textwidth]{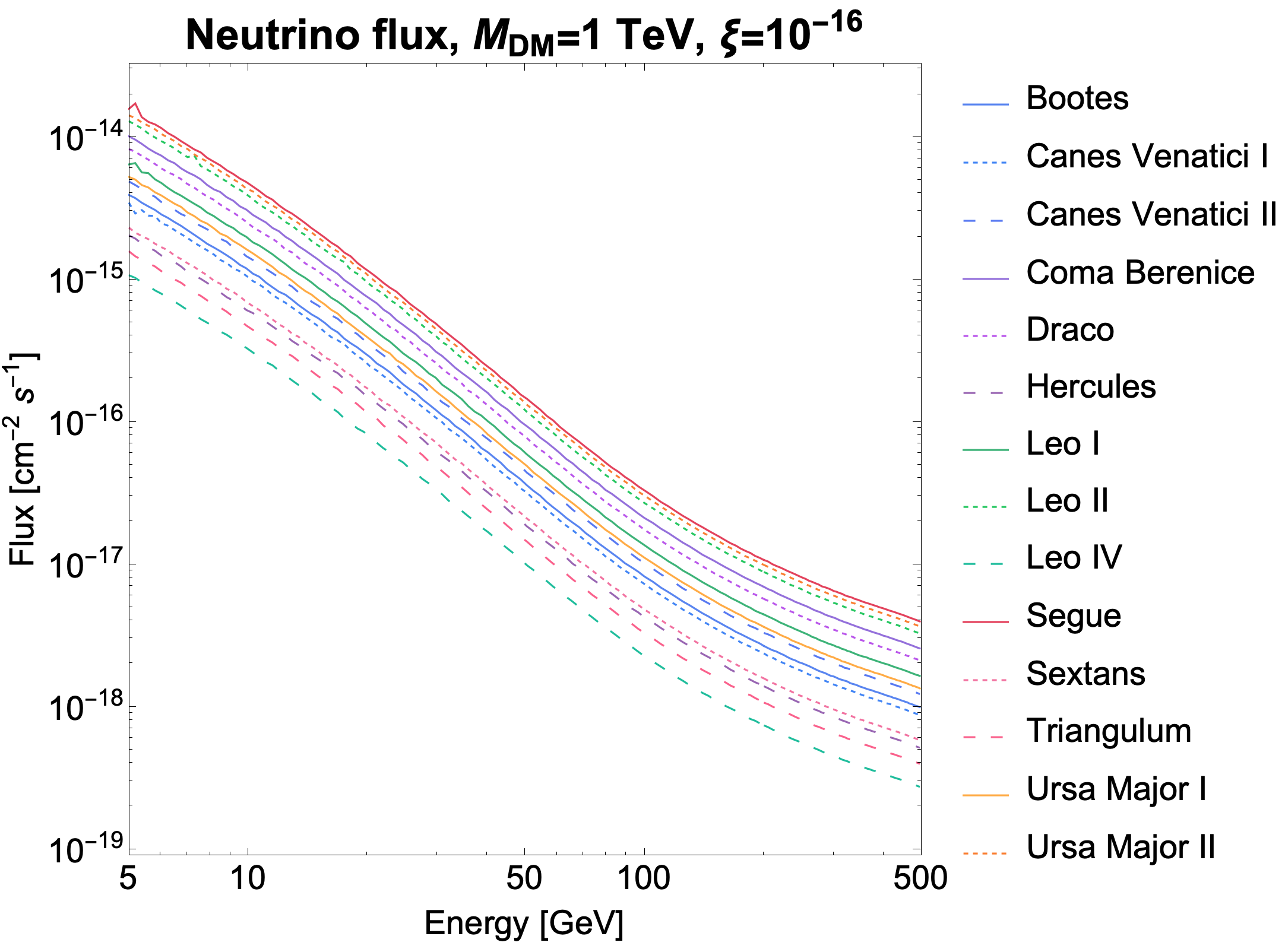}
        \subcaption{}
    \end{minipage}
    \caption{Flux of neutrinos in Earth produced by decaying DM in the dSphs, where neutrino oscillation is considered. Each line corresponds to one dSph. For each plot there is a fixed value for the mass of the candidate and the coupling parameter. In the first row $\xi=10^{-10}$, in the second is $10^{-13}$ and for the third one is $10^{-16}$. And for each column the mass is 10 GeV in first one, 100 GeV in the second one and 1 TeV in the last one.}
    \label{fig:Nu-flux-dSph}
    \vspace{0.5em}
\end{figure}

For the neutrinos, we also considered the oscillation following the same procedure as the one described in the section of the Milky Way. The results are shown in Figure ~\ref{fig:Nu-flux-dSph}.

The ratio between the fluxes in the first and second columns is approximately $10^{5}$, and between the second and third columns is only 6. The total flux is the sum of the flux of muon neutrinos that arrive in Earth from the source. 

\begin{figure}
    \centering
    \begin{minipage}[b]{0.295\textwidth}
        \centering
        \includegraphics[width=\textwidth]{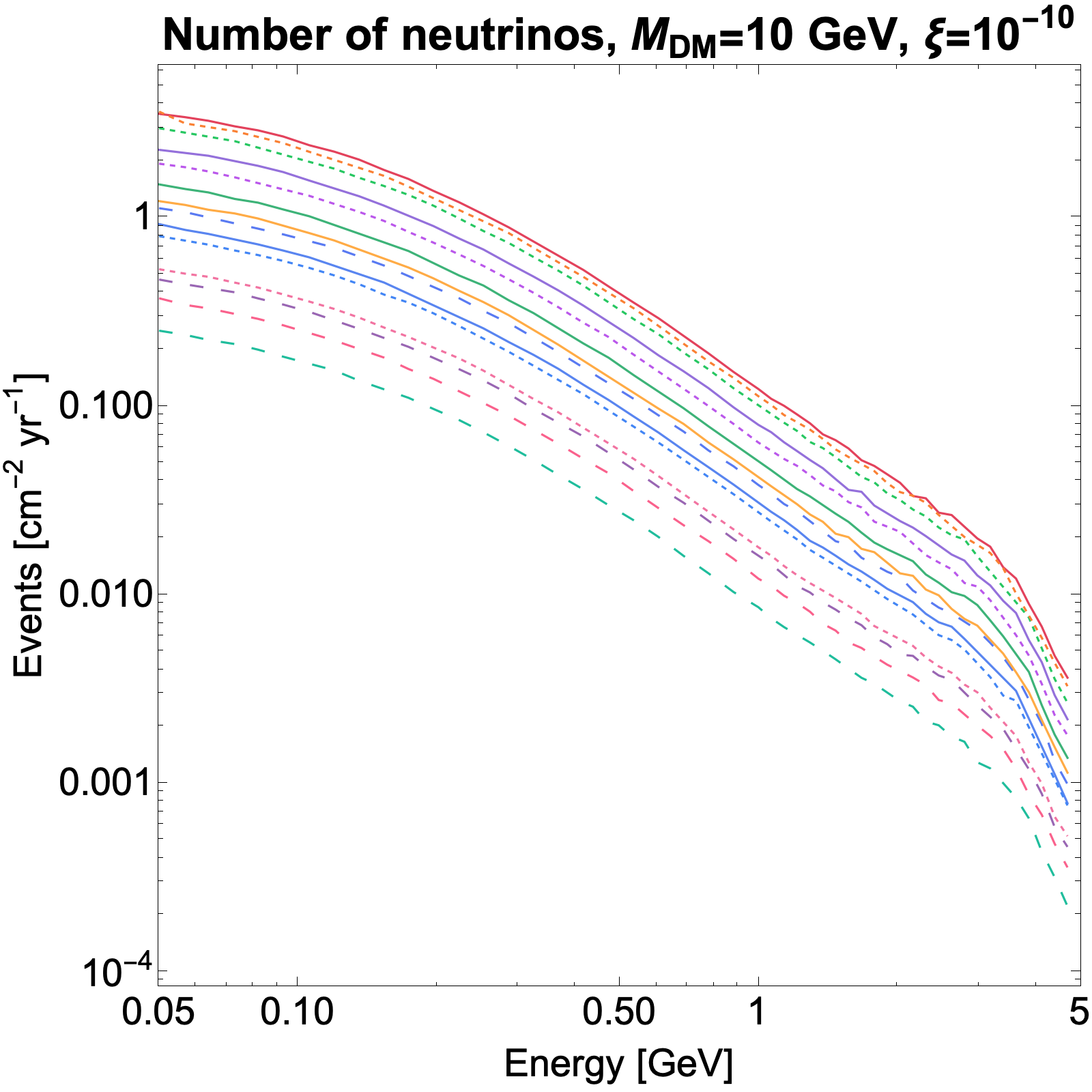}
        \subcaption{}
    \end{minipage}
    \begin{minipage}[b]{0.295\textwidth}
        \centering
        \includegraphics[width=\textwidth]{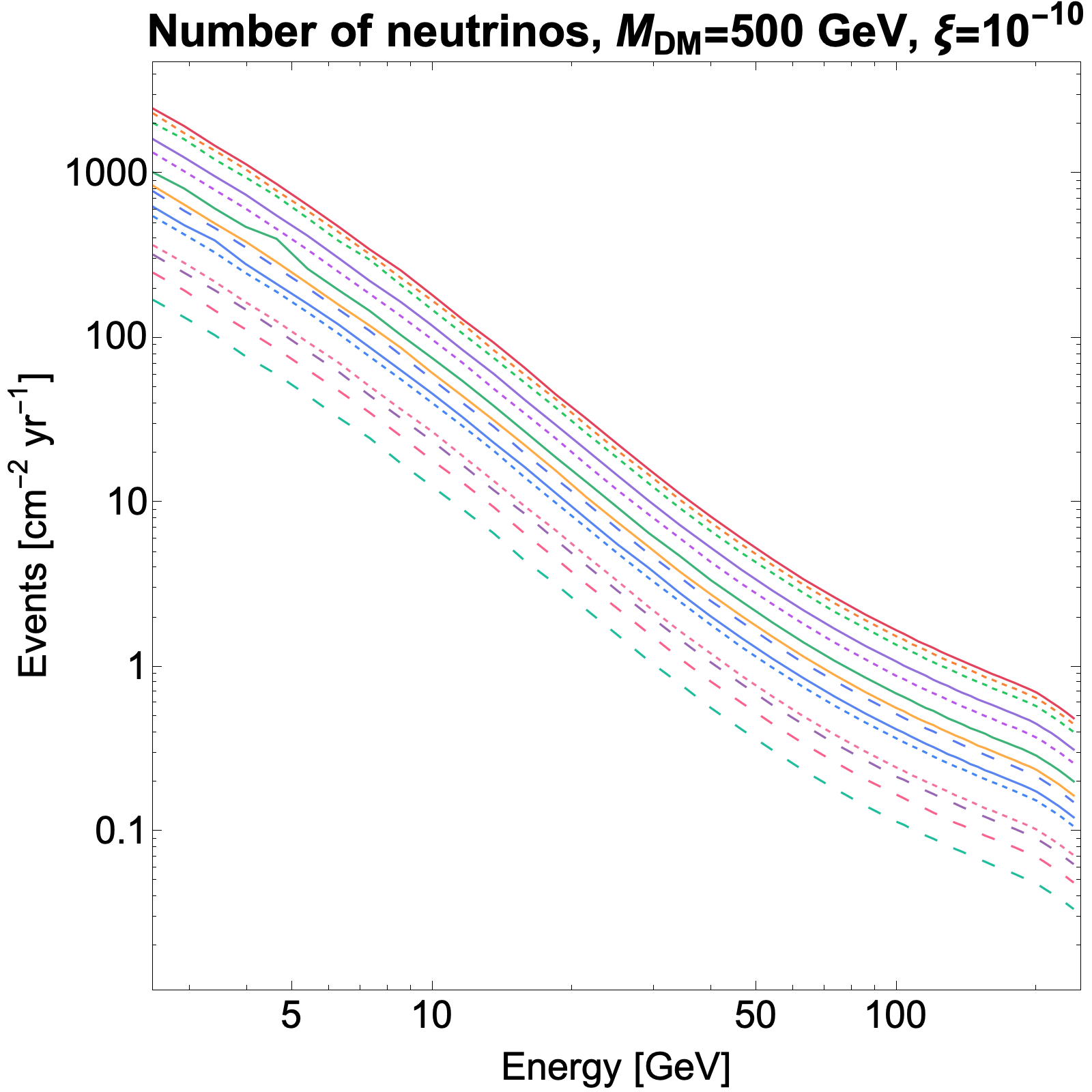}
        \subcaption{}
    \end{minipage}
    \begin{minipage}[b]{0.39\textwidth}
        \centering
        \includegraphics[width=\textwidth]{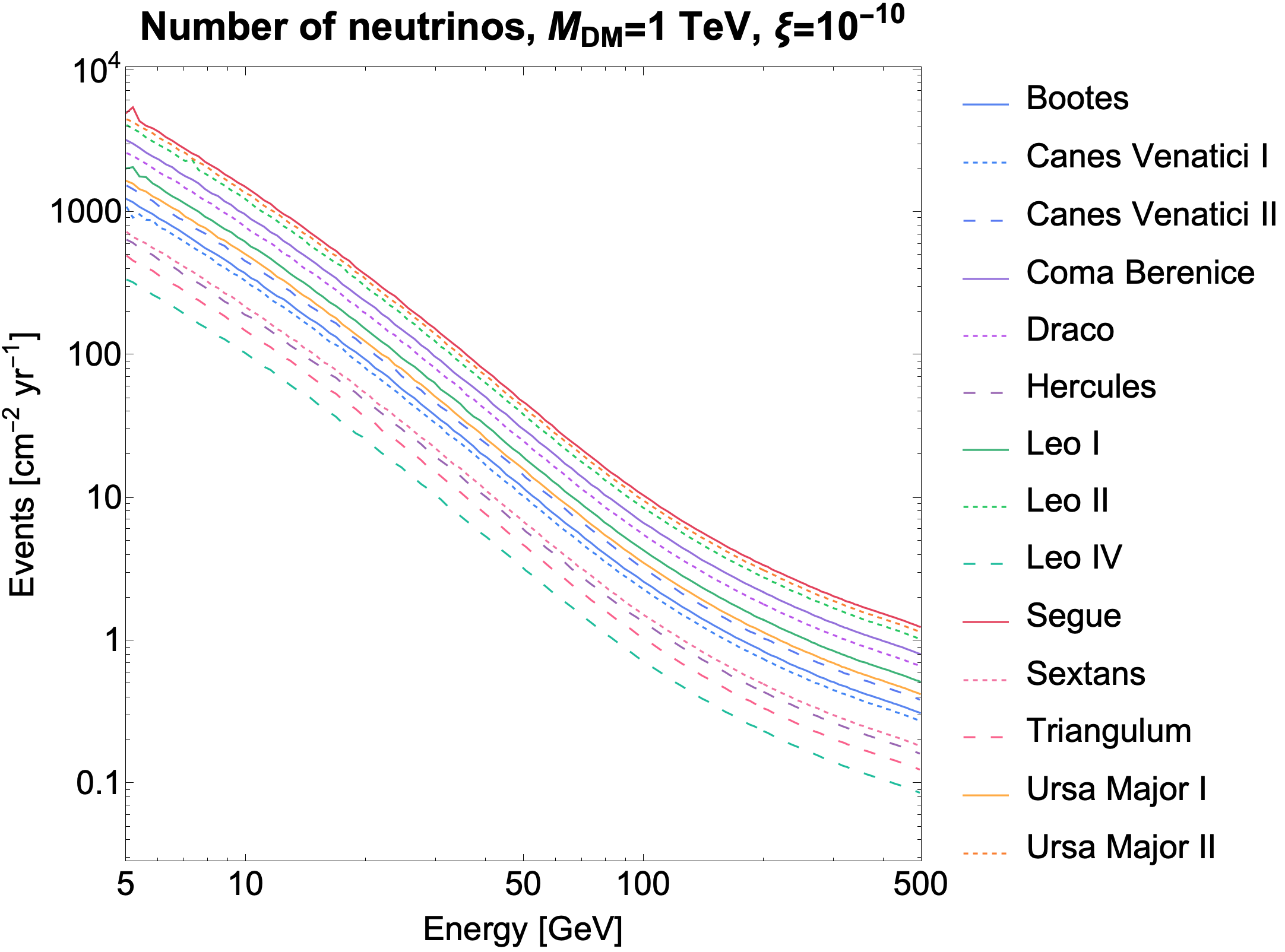}
        \subcaption{}
    \end{minipage}
    \vspace{0.5em}
    \begin{minipage}[b]{0.295\textwidth}
        \centering
        \includegraphics[width=\textwidth]{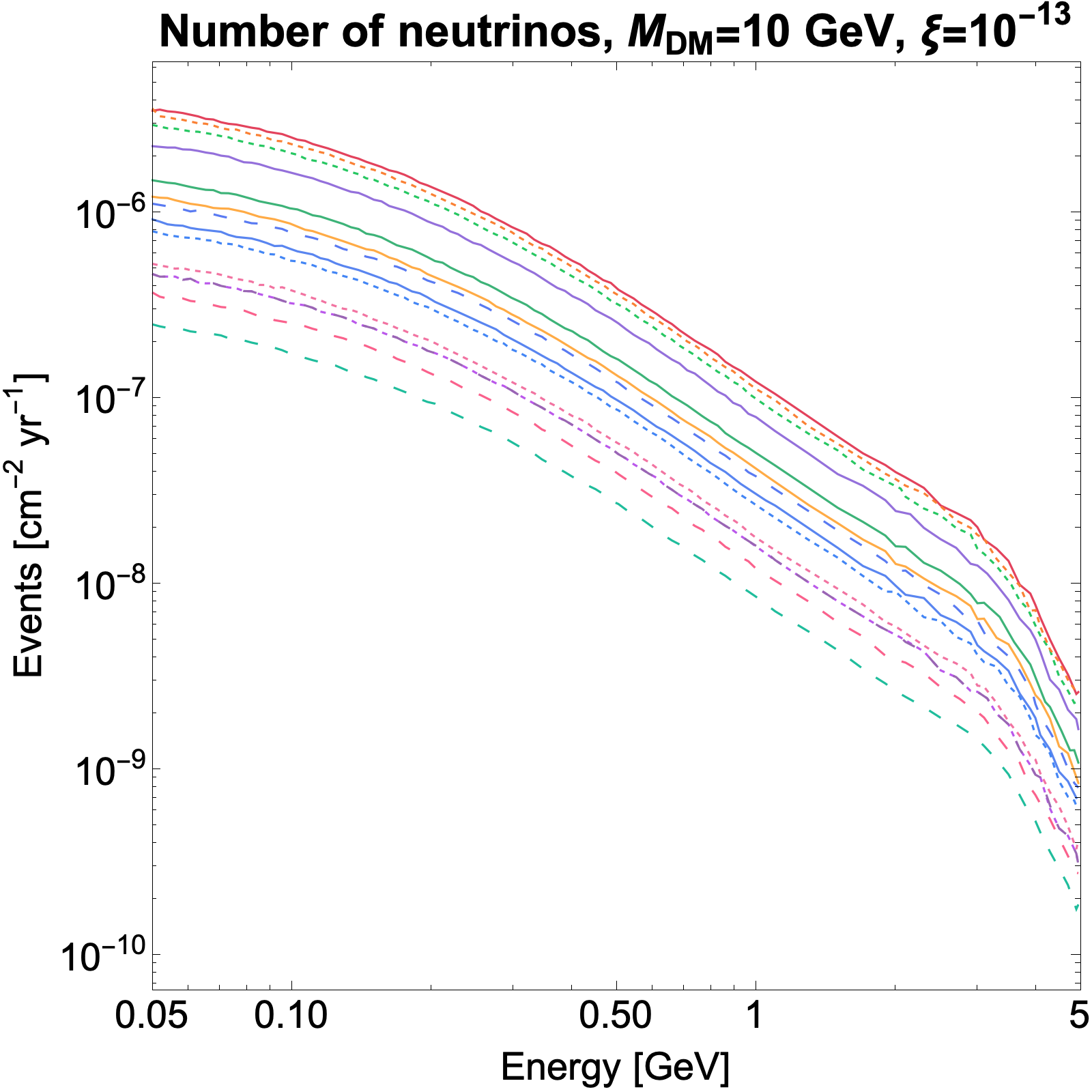}
        \subcaption{}
    \end{minipage}
    \begin{minipage}[b]{0.295\textwidth}
        \centering
        \includegraphics[width=\textwidth]{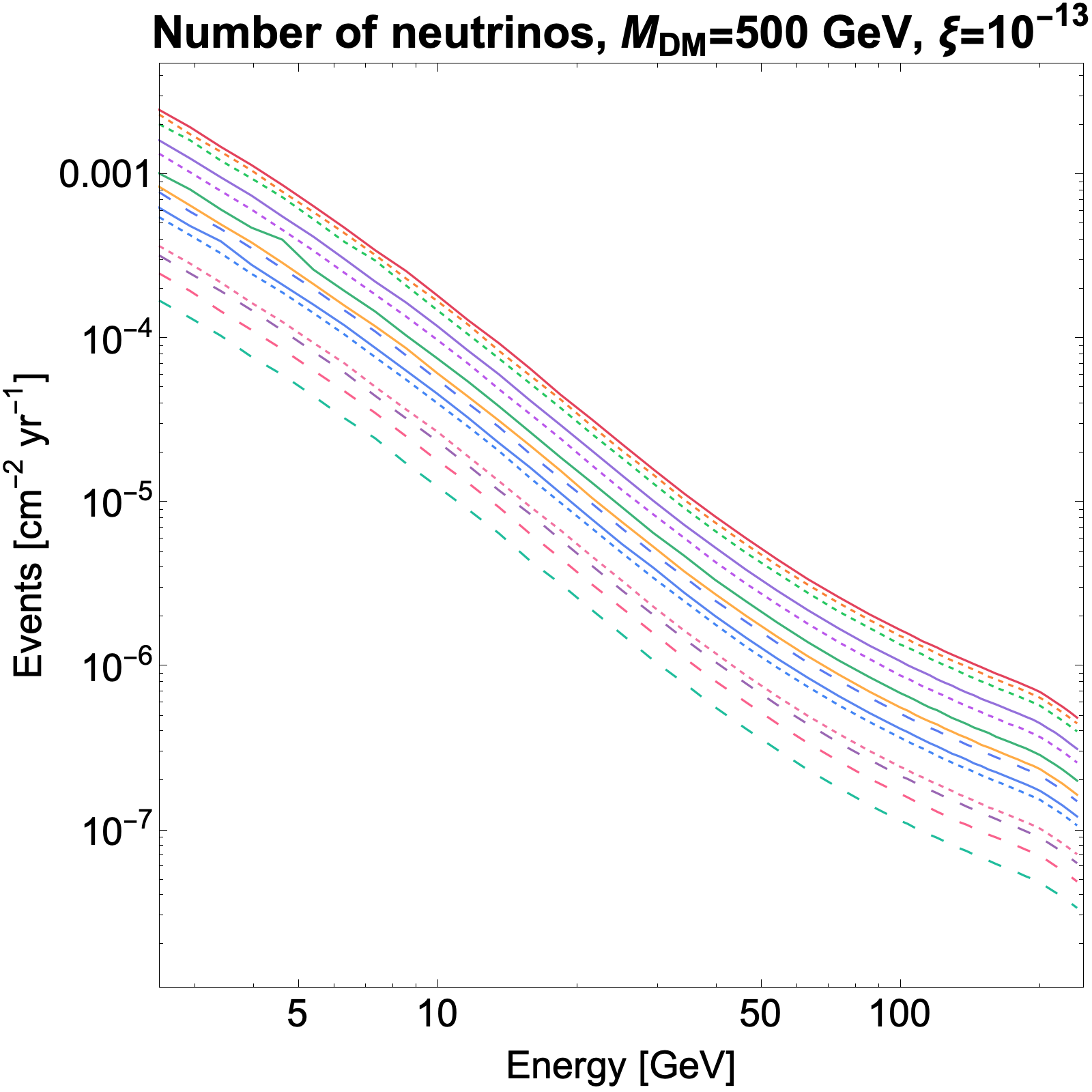}
        \subcaption{}
    \end{minipage}
    \begin{minipage}[b]{0.39\textwidth}
        \centering
        \includegraphics[width=\textwidth]{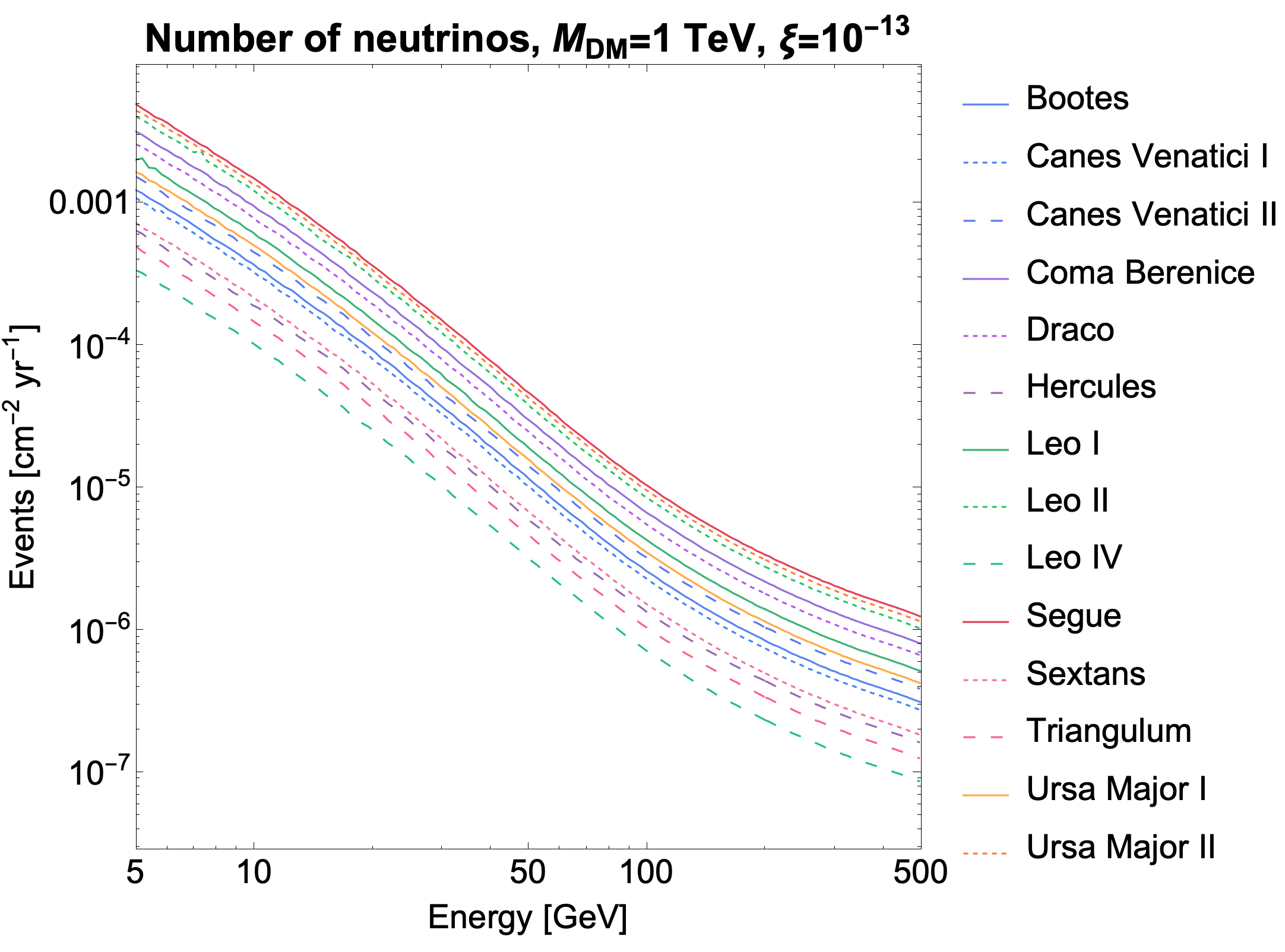}
        \subcaption{}
    \end{minipage}
    \vspace{0.5em}
    \begin{minipage}[b]{0.295\textwidth}
        \centering
        \includegraphics[width=\textwidth]{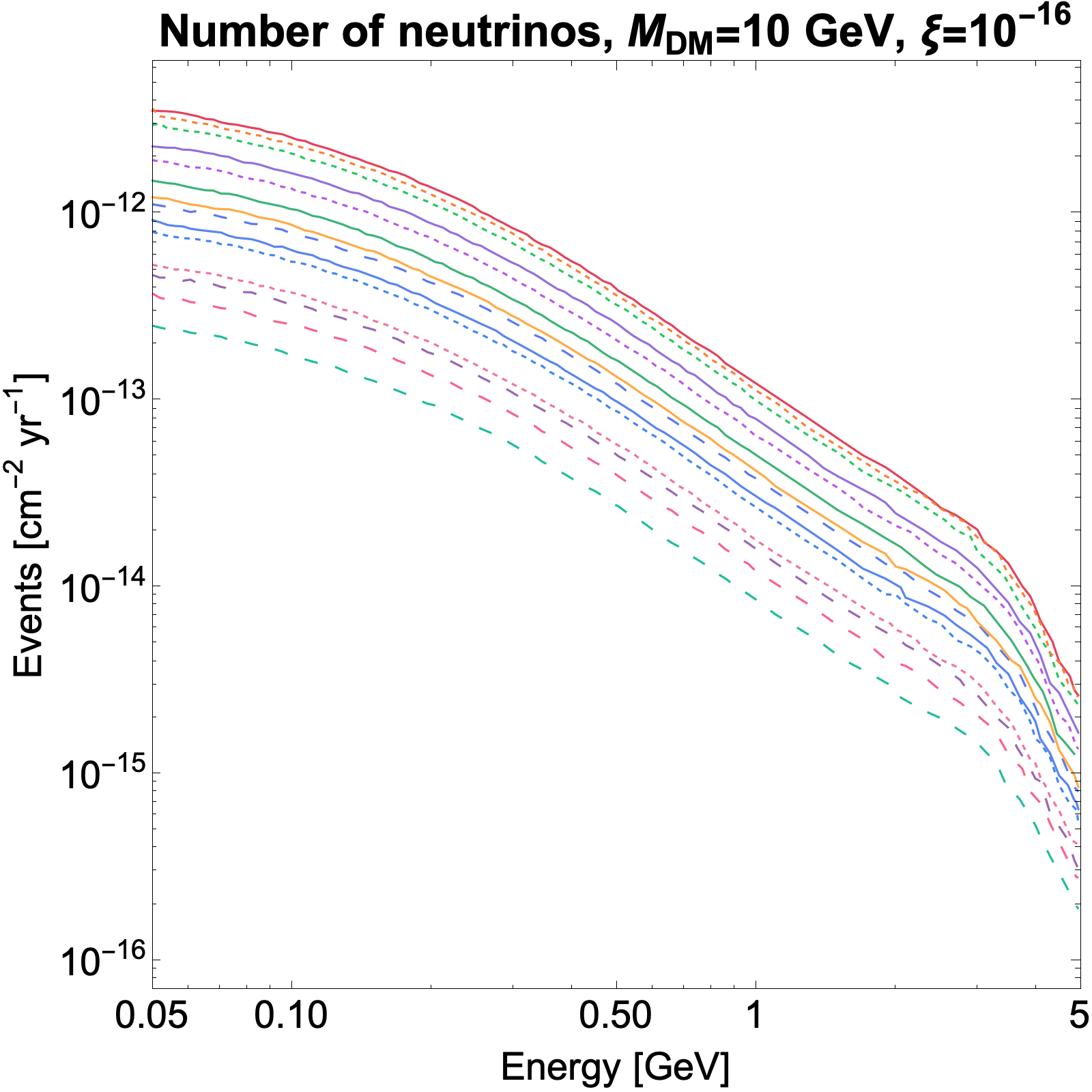}
        \subcaption{}
    \end{minipage}
    \begin{minipage}[b]{0.295\textwidth}
        \centering
        \includegraphics[width=\textwidth]{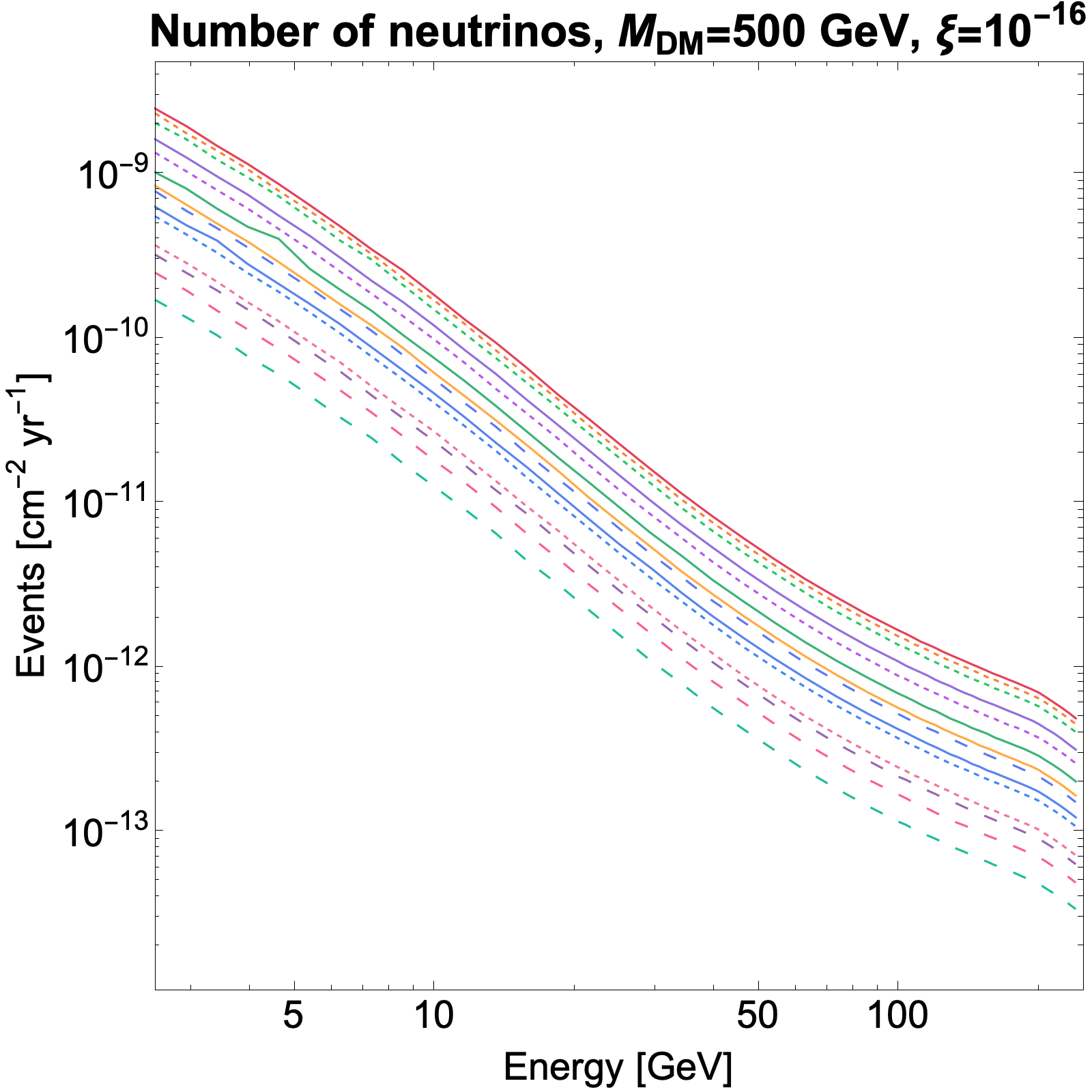}
        \subcaption{}
    \end{minipage}
    \begin{minipage}[b]{0.39\textwidth}
        \centering
        \includegraphics[width=\textwidth]{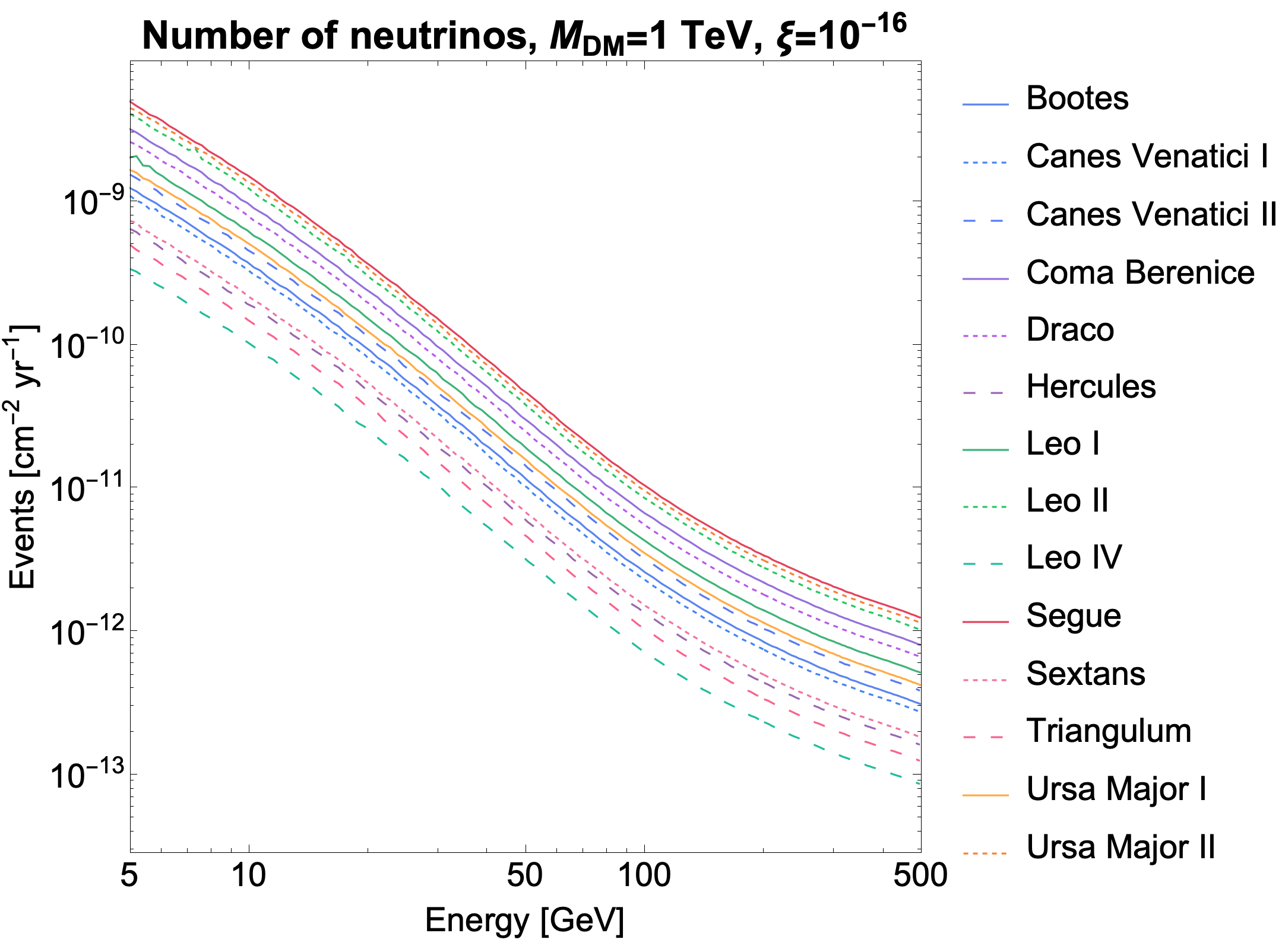}
        \subcaption{}
    \end{minipage}
    \caption{Number of neutrinos in the Earth, taking on count the phenomen of oscillation. Each line corresponds to one dSph. For each graph there is a fixed value for the mass of the candidate and the coupling parameter. In the first row $\xi=10^{-10}$, in the second is $10^{-13}$ and for the third one is $10^{-16}$. And for each column the mass is 10 GeV in first one, 100 GeV in the second one and 1 TeV in the last one.}
    \label{fig:Nu-Eve-dSphs}
\end{figure}

Finally, we estimated the number of muon neutrinos that can be detected with a detector located on Earth, considering an effective area of one squared kilometer and one year data. The results are shown in Figure ~\ref{fig:Nu-Eve-dSphs}. 

We can observe that for the DM mass of 1 TeV and the coupling parameter $10^{-10}$ the number of neutrinos with energy in the range of 5 to 10 GeV is three hundred to three thousand. In the case for DM mass of 10 GeV between 2 and 5 GeV we can see some noise. For the most energetic neutrinos we could need at least 10 years of data to detect one neutrino. Also, the dSphs that produce the most neutrinos are Segue I, Ursa Major II and Leo II, as their DM density are the highest of all the objects in this study. 

\section{Conclusions}
\label{sec:conclusions}

In this work, we have explored an indirect detection scenario for decaying dark matter (DM) within a high-energy physics framework. Models with unstable DM require lifetimes significantly larger than the age of the Universe. Current bounds constrain the lifetime to be $\gtrsim 10^{24}\,\text{s}$, and the scenario considered here is consistent with these limits. The DM mass range analyzed, from $10$ GeV to $1$ TeV, lies within the reach of present and future experiments. Neutrino telescopes are particularly sensitive to lower masses ($10$--$500$ GeV), whereas gamma-ray observatories are more relevant in the TeV regime, although $1$ TeV often lies close to their lower energy thresholds.

The gamma-ray and neutrino fluxes computed for the Milky Way halo and dwarf spheroidal galaxies (dSphs) are of comparable order of magnitude. This is consistent with their similar D-factors and DM densities. Although some predicted fluxes fall below current sensitivities, others are within the reach of indirect detection experiments. The most favorable scenarios correspond to shorter lifetimes ($\xi = 10^{-10}$) and DM masses of $500$ GeV and $1$ TeV. In all cases, the signal is strongly concentrated toward the central regions of the halos, which is consistent with the expectations from DM density profiles.

To estimate the number of detectable events, we considered an idealized detector with a large effective area and no dependence on the arrival direction. Under these assumptions, the predicted number of events exceeds current observations, indicating that future experiments could be sensitive to the scenarios studied here. In particular, next-generation neutrino telescopes such as KM3NeT may increase the number of detectable events by up to $46\%$.

A comparison between gamma-rays and neutrino signals shows no clear preference for either channel, as both fluxes are of similar magnitude. The energy spectra exhibit a decreasing behavior with energy, implying that lower-energy photons and neutrinos are more likely to be detected. This trend is also reflected in the expected event rates.

Current experimental constraints on decaying DM scenarios remain limited. The most stringent bounds apply to the DM lifetime, while other model parameters are largely unconstrained. A relevant limitation arises from the computational tools employed, which are typically restricted to two-body decay channels. This constrains the accessible parameter space, particularly at higher masses where multi-body decays may become important.

Despite extensive efforts, none of the main DM search strategies—direct detection, collider production, and indirect detection—has provided conclusive evidence so far. Among them, indirect detection remains a particularly promising approach, as it probes astrophysical environments without relying on direct DM-SM interactions or collider energy reach. In this context, decaying DM scenarios are especially well motivated within the broad landscape of viable models. Although only a limited number of experiments currently target this mechanism, future experimental improvements and planned facilities are expected to significantly enhance the sensitivity to such scenarios and further constrain the available parameter space, ultimately contributing to a deeper understanding of the nature of dark matter and its role in the Universe.
\acknowledgments

The authors L.~L.-L. and A.~C.-M. are supported by SECIHTI through the SNII program. F.~S.~J.-V. is supported by SECIHTI through the National Graduate Scholarship Program.

\bibliographystyle{JHEP}
\bibliography{referenceMatrix.bib}

\end{document}